\documentclass[10pt,a4paper,twocolumn]{article}
\usepackage[utf8]{inputenc}

\usepackage[english]{babel}
\usepackage{blindtext}

\usepackage{multirow}
\usepackage{multicol}
\usepackage{soul}
\usepackage{xcolor}
\usepackage{subcaption}
\usepackage{graphicx}
\usepackage{url}
\usepackage[numbers,compress]{natbib}
\usepackage{hyperref}

\usepackage[margin=0.75in]{geometry}

\author{Jordi Paillisse$^{\dag\ddag}$, Marc Portoles$^{\ddag}$, Albert Lopez$^{\dag}$, Alberto Rodriguez-Natal$^{\ddag}$, David Iacobacci$^{\S}$\thanks{Work performed while at Cisco}, \\
Johnson Leong$^{\P}$\footnotemark[1],  Victor Moreno$^{\ddag}$, Albert Cabellos$^{\dag}$, Fabio Maino$^{\ddag}$ and Sanjay Hooda$^{\ddag}$ \\
$^{\dag}$\textit{\small UPC-BarcelonaTech, Barcelona, Spain} - \textit{\small \{jordip, alopez, acabello\}@ac.upc.edu,} \\
$^{\S}$\textit{\small BMP LLP, CA, USA}, $^{\P}$\textit{\small Uber Technologies Inc., San Francisco, CA, USA}, \\
$^{\ddag}$\textit{\small Cisco, San Jose, CA, USA - \{mportole,natal,vimoreno,fmaino,shooda\}@cisco.com}}

\title{\textbf{SD-Access: Practical Experiences in Designing and Deploying Software Defined Enterprise Networks}}

\begin{document}

\maketitle

\begin{abstract}

Enterprise Networks, over the years, have become more and more complex trying to keep up with new requirements that challenge traditional solutions. Just to mention one out of many possible examples, technologies such as Virtual LANs (VLANs) struggle to address the scalability and operational requirements introduced by Internet of Things (IoT) use cases. To keep up with these challenges we have identified four main requirements that are common across modern enterprise networks: (i) scalable mobility, (ii) endpoint segmentation, (iii) simplified administration, and (iv) resource optimization. To address these challenges we designed SDA (Software Defined Access), a solution for modern enterprise networks that leverages Software-Defined Networking (SDN) and other state of the art techniques. In this paper we present the design, implementation and evaluation of SDA. Specifically, SDA: (i) leverages a combination of an overlay approach with an event-driven protocol (LISP) to dynamically adapt to traffic and mobility patterns while preserving resources, and (ii) enforces dynamic endpoint groups for scalable segmentation with low operational burden. We present our experience with deploying SDA in two real-life scenarios: an enterprise campus, and a large warehouse with mobile robots. Our evaluation shows that SDA, when compared with traditional enterprise networks, can (i) reduce overall data plane forwarding state up to 70\% thanks to a reactive protocol using a centralized routing server, and (ii) reduce by an order of magnitude the handover delays in scenarios of massive mobility with respect to other approaches. Finally, we discuss lessons learned while deploying and operating SDA, and possible optimizations regarding the use of an event-driven protocol and group-based segmentation.

\end{abstract}

\section{Introduction}

Current Enterprise Networks (EN) present a high degree of complexity derived from their organic evolution. Traditional ENs are built in a three-tier structure: access, distribution and core, with each layer leveraging a distinct set of protocols. The access-distribution segment uses L2 technologies such as Virtual Local Area Networks (VLANs), Spanning Tree Protocol (STP), and VLAN Access Control Lists (ACLs). On the other hand, the distribution-core segment is usually L3 with a combination of Open Shortest Path First (OSPF), Multiprotocol Label Switching with Label Distribution Protocol (MPLS-LDP), and IP ACLs. This design varies from  deployment to deployment, but the end result are complex networks that can be running up to tens of different protocols across hundreds of switches. This makes it hard to adapt to new requirements, such as isolating  IoT devices or stretching L2 domains across distributed locations. Even if some of these challenges might have been addressed by technologies designed for service providers, such as Virtual Routing and Forwarding (VRF), given the high port density required in typical campus networks, adopting the same technologies in enterprise solutions is not cost effective.

Specifically, current ENs lack three key elements. First, scalable endpoint mobility across all the enterprise facilities, to address the ever-increasing amount of roaming devices. Usually this is handled via sending all wireless traffic through centralized Wireless LAN (WLAN) controllers, which limits scalability, and reduces bandwidth. Second, simple to operate segmentation. The most common forms of segmentation in ENs are VLANs or VRFs, which do not scale well and can be difficult to configure at scale. Another example are IP-based ACLs, that over time can easily become long and difficult to map to the original intent. Third, simplified operations. Network administrators might configure each router individually, and, as we mentioned previously, have to deal with a myriad of different protocols. Although there exist more modern solutions \cite{casado2007ethane, lebrun2018software} that satisfy some of these requirements, we believe that the state of the art does not deal with scale and dynamism in a cost-effective manner, i.e. without requiring large capital expenditures (CAPEX).

CAPEX plays, indeed, a very important role in the context of Enterprise Networks. First, because of brownfield deployments:  typically, network administrators do not want to upgrade \emph{all} of their switches for new features. Since these are usually legacy devices with limited features, new network designs require a way to add new functionality without a forklift upgrade. Second, because deploying devices with reduced FIB size or CPU power decreases CAPEX but in turn means less powerful devices that require resource optimization.

In this paper we present the rationale, implementation, evaluation and experience matured in deploying SDA (Software Defined Access). Our objective with SDA was to design a solution that addressed the aforementioned requirements of modern EN. With this goal in mind, we leveraged a vast spectrum of research ideas, architectures and protocols produced by the community in the last decade \cite{Koponen2014, dbSDNcontroller, Casado:2012:FRE:2342441.2342459, smartMPLSvpn, gbpWhitepaper, eventDrivenProtocol}, and integrated them in a practical and deployable solution. First, we leveraged network overlays as discussed in the Fabric architecture \cite{Casado:2012:FRE:2342441.2342459}, in the form of the Locator/ID Separation Protocol (LISP \cite{ietf-lisp-rfc6833bis-27}). This offers three benefits: (i) we can upgrade existing deployments (brownfields) with minimal touch, (ii) their layer of indirection makes it easy to support L3 mobility, and (iii) they make segmentation with VRFs more scalable. Second, we applied the Software-Defined Networking (SDN) principle of centralized control to track endpoint location, and map endpoints to segmentation policies across the whole network. Finally, we chose a reactive protocol (LISP) to distribute network state to the data plane. In other words, we populate the switches  forwarding tables only if required by the active traffic pattern. This reduces the overall switch requirements in terms of FIB size and CPU power which results in reduced CAPEX.

We must remark that this paper is not meant to introduce a novel architecture, but rather an account of our experience and lessons learned in designing, and deploying modern enterprise networks.  We describe how we combine state of the art techniques to realize a practical solution, and the lessons that we have learnt when operating SDA. We detail our experience through two real-life deployments. First, a medium-sized enterprise campus network serving around 450 endpoints that includes fixed hosts, mobile hosts, application servers, IoT devices, etc. We show that in this scenario, our reactive protocol optimizes data plane state with a 70\% reduction of FIB entries in the data plane compared to solutions that store all the state in all routers (e.g. BGP). Second, a large warehouse where hundreds of robots are moving at speed to fulfill shipping orders, such as those ran by large on-line retailers. Here we evaluate the handover delay of 16,000 robots triggering 800 mobility events per second. Our solution achieves 5 times lower handover delay compared to existing approaches. Finally, we present our lessons learned from deploying SDA, such as reducing the initial connection delay due to the reactive protocol, coping with connectivity issues in the underlay or dealing with policy updates at scale.

\section{Requirements and Design Decisions}

\begin{table*}[!tp]
\small
\caption{Summary of current state of the art, challenges, and design decisions}
\begin{center}
\begin{tabular}{|l|c|c|c|c|c|c|}
\hline
\textbf{Requirement} & \textbf{Current Approach}    & \textbf{Limitations} &  \textbf{Our approach} &\textbf{Benefits}  \\
\hline
\multirow{2}{*}{Resource efficiency} & BGP and OSPF         & \multirow{2}{*}{Granularity}  & Traffic-driven &   Reduced CAPEX,\\
                                     & prefix aggregation  &             & route learning &  device-level granularity\\
\hline

\multirow{2}{*}{Mobility} &  L2 centralized   & Scalability, &   L3 centralized control, & Increased scalability,\\
                          &  control and data plane  & triangular routing  & distributed data plane &  optimized routing  \\
\hline
\multirow{2}{*}{Segmentation} & \multirow{2}{*}{VLANs and VRFs}   & Scalability,  & Limited L2 stretching, & Increased scale  \\
                            &                 & network-wide policies  & 'centralized' VRFs    & with less resources\\
\hline
Simplified  & VLAN and   &  Error-prone,  & \multirow{2}{*}{Group-based policies} &  Smaller ACLs, end-  \\
                  administration                 & IP-based ACLs   &  no mobility   &  &  to-end enforcement \\

\hline
\end{tabular}
\label{tab:decision}
\end{center}
\end{table*}

This section delves deeper into the requirements of current ENs, explains limitations in the state of the art and discusses design decisions, summarized in table \ref{tab:decision}.

\textbf{Resource optimization}: Routing protocols commonly found in enterprises, such as OSPF, IS-IS or BGP make it difficult to reduce the FIB space without losing granularity\footnote{It is usual to leverage BGP prefix aggregation or OSPF Areas to reduce the overall number of routes in the network, at the price of less granularity.}. 
We approached this challenge by leveraging a reactive protocol (LISP in our implementation), rather than a proactive. Instead of pushing all routing entries beforehand to the routers, we only retrieve the necessary forwarding entries from a centralized server on-demand, and only for the routers that need them \cite{reactiveAndProactive, eventDrivenProtocol}. We track endpoints by their IP address, so that routers download routes for the remote endpoints they need to reach, based on incoming traffic from local endpoints.

In addition, this reactive approach is helpful with mobility because we can reduce convergence time. This arises from the fact that a reactive approach  reduces the churn generated by the location updates: we only notify the parties affected by a specific mobility event. We must remark that a reactive protocol presents several challenges, such as a potential initial delay for the establishment of flows (sec. \ref{sec:sec:routing}), or detecting connectivity outages in the underlay (sec. \ref{sec:lessons} discusses our learnings in this space).

\textbf{Mobility}: the current trend of wireless first makes it critical to support a large amount of wireless endpoints. Traditionally, ENs handle mobility at L2 in a centralized way for both  data plane and control plane. A gateway device (WLAN controller) acts as a sink for all traffic from all access points, performs access control, and re-injects it to the L3 network. This approach presents a serious scalability limitation because the gateway device becomes a bottleneck \footnote{Although this particular concern can be alleviated with hierarchical controllers, it comes at the price of increased complexity and number of devices.}. In addition, it creates triangular routing because all L3 traffic is forced to go to the gateway and then  back to the actual destination.

SDA tackles mobility at layer 3 using network overlays \cite{rfc3344mobileIPv4, mobilityOverlays}, specifically with the mobility features of LISP. We keep the wireless control plane centralized for authentication purposes, but we let packets coming from the access point to be directly routed to destination. This distributed data plane greatly increases scalability. 

\textbf{Segmentation}: traditionally, the most common form of segmentation in enterprise networks are Virtual Local Area Networks (VLANs \cite{vlanSTD2018}) and Virtual Routing and Forwarding tables (VRFs \cite{rfc4364}). Despite their simplicity, VLANs scope must be kept limited  to prevent flooding of broadcast traffic  or L2 forwarding loops, hence, they do not scale well. Regarding VRFs, they scale better than VLANs, but since each device has to be individually configured it is hard to implement global polices across the whole network. A direct consequence is that administration becomes too cumbersome as the number of VRFs increases. In addition, both of these approaches present similar limitations that make it hard to deal with mobility at scale. 

SDA addresses these issues at different levels: for L2 segmentation \cite{Koponen2014, overlayL2stretch}, we carefully stretch L2 domains (c.f. sec. \ref{sec:sec:l2}). For L3, we still use VRFs, but map local VRFs to global virtual networks in order to handle L3 segmentation at scale. This way, network administrators only have to specify the virtual network for each endpoint \cite{smartMPLSvpn}. Finally, we add a layer of indirection to ease administration, detailed in the next paragraph. 

\textbf{Simplified administration}: a direct consequence of using network primitives such as IP addresses or VLANs for segmentation and access control is operational complexity \cite{aclUpdate}. In other words, network administrators have to translate business intent into IP addresses and ACLs and backwards. In the long run, this approach does not scale, is error prone, and increases complexity. 

To overcome this problem, we make use of the well-established group-based paradigm \cite{gbpWhitepaper, stone2001network} to define ACLs between groups, instead of IP prefixes. First, the network operator defines a connectivity matrix among all groups. Then it adds endpoints to each group. On the network level, routers track each endpoint by its IP address and add a 16-bit tag representing its group, so they can enforce the connectivity rules in the matrix. The benefit is that network administration is radically simplified and common operations, such as IP address planning or ACL configurations can be automated.

\section{Design and Implementation}

In this section, we describe the design and implementation details of SDA. First we provide an overview, then we describe the control plane and data plane. Finally, we detail how we support endpoint mobility and L2 services.
\subsection{Overview}

\begin{figure*}[!tp]
\centering
\includegraphics[width = 0.9\textwidth, keepaspectratio  ,trim={0cm, 0cm, 0cm, 7cm}, clip]{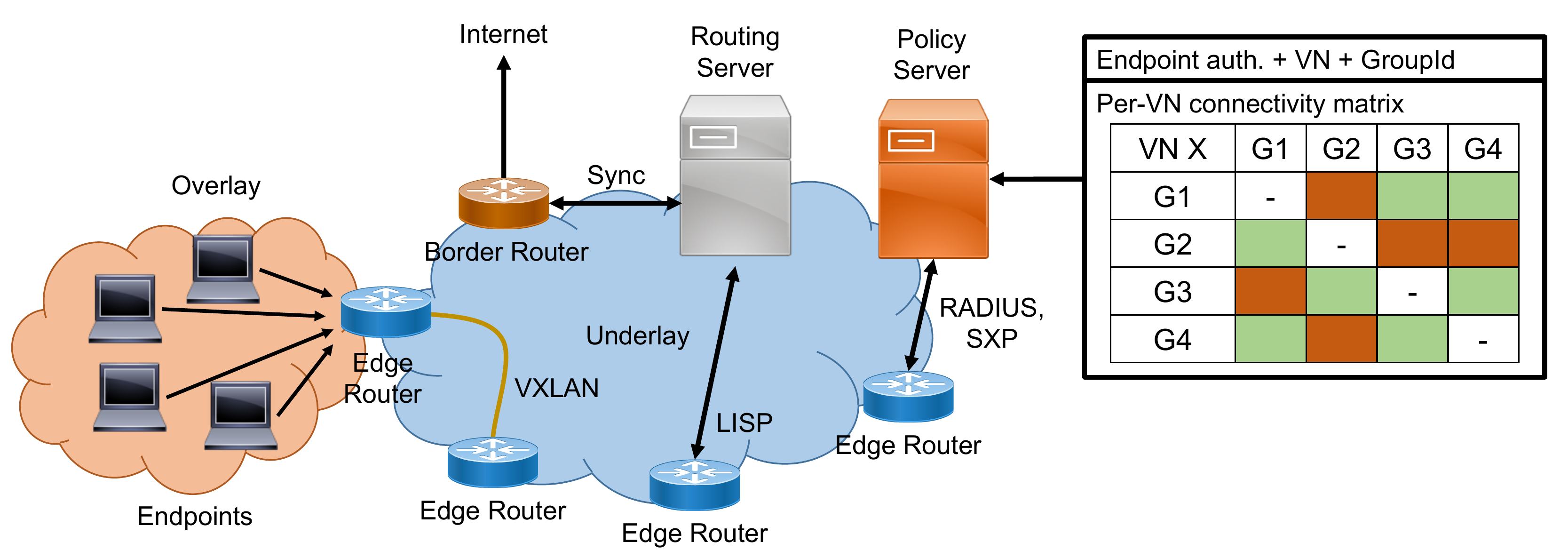}
\caption{Global design }
\label{fig:arch}
\end{figure*}

On a conceptual level our design presents the usual data/control plane layer separation typical of SDN \cite{mckeown2008openflow}. Figure \ref{fig:arch} presents an overview of the design. We expose a simple declarative interface for network operators to define: (i) an endpoint's group and the Virtual Network (VN) where they can connect, (ii) the endpoint authentication data, and (iii) the connectivity matrix across groups of endpoints.

We store this information in the control plane in two different servers, a routing server and a policy server. The policy server authenticates endpoints, assigns them a group and configures the data plane routers with the required group rules from the connectivity matrix. The routing server keeps track of all endpoints by their IP address and provides routes upon demand by the data plane. 

On the data plane, the overlay routers - hereafter referred to as \emph{edge} routers - enforce the connectivity matrix and route packets to the corresponding edge router. A special \emph{border} router provides access to external networks. Finally, endpoints can roam freely across edge routers.

\subsection{Control Plane}
We based the control plane design on a database-focused approach similar to \cite{dbSDNcontroller}, as opposed to more traditional designs \cite{gude2008nox, koponen2010onix, pankaj2014onos}. The control plane consists of two logically centralized servers: policy server and routing server\footnote{The reason for this separation is that usually the host onboarding process needs both the endpoint credentials and group permissions, while the normal packet flow only needs the location mappings.}. Table  \ref{tab:mappings} presents a summary of all our control plane data. 

\subsubsection{Control Plane Policy Sever}
We offer two degrees of segmentation: Virtual Networks (VN) and the group connectivity matrix. These group rules are independent for each VN. On one hand, VNs offer strong isolation at a 'macro' level. An example is a hospital network, where we want to isolate the doctors, guest and medical devices networks, we never expect them to be able to communicate each other. This is especially relevant to isolate legacy devices susceptible to attacks, e.g. an MRI machine running an outdated OS. In addition, it is a way to mitigate lateral spread attacks. 

On the other hand, the group rules offer a 'micro' segmentation for  finer grain control inside a VN. For example, this level of segmentation can separate different types of devices within a VN in Bring Your Own Device scenarios.

The policy server stores the connectivity rules from the connectivity matrix, and, for each endpoint: its authentication data, and associated GroupId and VN. GroupId and VN are 16-bit and 24-bit identifiers, respectively. The authentication data is variable since we support different  RADIUS-based authentication protocols \cite{rfc2865}, both with EAP or without. We use a specific protocol, Scalable-Group Tag eXchange Protocol (SXP \cite{sxp}) to distribute the GroupIds and connectivity rules to edge routers. From the network perspective, VNs are mapped to isolated routing-switching domains, while GroupIds are inputs to group-based ACLs.

\subsubsection{Control Plane Routing Server}\label{sec:sec:routing}
The routing server stores the endpoint location, i.e. pairs of overlay-to-underlay IP addresses plus its associated VN. The overlay IP is the IP used by endpoints, while the underlay IP is the IP of the edge router serving this endpoint.  The other edge routers encapsulate traffic for such endpoint towards the underlay IP. After a successful device onboarding (sec. \ref{sec:sec:onboarding}), or upon detecting a mobility event, edge routers update the underlay location of an overlay IP address. Edge routers also retrieve this mapping when they receive a connection request to a particular device. We leverage the control plane aspects of the Locator/ID Separation Protocol (LISP, \cite{ietf-lisp-rfc6833bis-27}).

In a nutshell, the LISP control plane offers two messages: Map Request, to retrieve the underlay address of an overlay endpoint, and Map Register, to update the location of an endpoint, i.e. the overlay to underlay mapping. This way, we can store in the data plane the overlay-to-underlay mappings that are required by the edge router to serve  incoming traffic. In addition, the LISP control plane supports mobility, is well suited for SDN architectures \cite{rodriguez2015lisp}, and accommodates different overlay address families apart from IP, e.g MAC addresses. This is especially helpful to support L2 services (sec. \ref{sec:sec:l2}).

\begin{table*}[!tp]
\small
\caption{Types of Control Plane Data}
\begin{center}
\begin{tabular}{|c|c|c|c|c|}
\hline
\textbf{Name} & \textbf{Key}    & \textbf{Value} &  \textbf{Updated by} \\
\hline
Endpoint Data &  Endpoint authentication info & GroupID , VN &  Network Operator\\
\hline
Group Rules & Source GroupId + Dest. GroupId  & Action (allow, deny) &  Network Operator\\
\hline
\multirow{2}{*}{Endpoint Location} &  \multirow{2}{*}{VN + Overlay IP addr.}  & Underlay IP addr. & \multirow{2}{*}{Edge Routers}\\
 &  & (i.e. edge router) & \\
\hline
\end{tabular}
\label{tab:mappings}
\end{center}
\end{table*}

However, a drawback of using a reactive protocol such as LISP is the initial packet loss until the edge router downloads the route for a new destination. We have overcome this issue by installing a default route in all edge routers that points to the border router, and by synchronizing the routing state in the border with the information in the routing server (sync arrow in fig. \ref{fig:arch}). This way, edge routers forward packets to the border until they finish the resolution process, during this time the border router forwards such packets to the destination. The border router is usually more powerful than edge routers in order to handle this extra load. Finally, some deployments may have more than one routing server or policy server for redundancy and load balancing.

\subsection{Data Plane}\label{sec:design:dataplane}
The data plane presents two distinct routers (fig. \ref{fig:arch}): edge and border. 

\textbf{Edge Routers:} perform four key functions. First, they encapsulate and decapsulate traffic from and to  endpoints, respectively. Second, provide inter-VN isolation ('macro' segmentation). We implement such segmentation with VRFs: LISP populates the VRF tables, and each entry has an associated GroupId. Third, they  detect roaming endpoints and update their location in the routing server. And fourth, enforce group permissions from the connectivity matrix ('micro' segmentation). 

\textbf{Border Routers:} perform the same functions as edge routers, with two exceptions. First, their FIB table is synchronized with the routing server. In other words, they don't use a reactive protocol, rather they are subscribed to all route updates from the routing server \cite{ietf-lisp-pubsub-05}. And second, they have routes to other networks, e.g. Internet, datacenter. Because of this, border routers have more powerful CPU and larger FIB tables. 

\begin{figure}[!tp]
\centering
\includegraphics[width=0.9\columnwidth, trim={2.25cm, 9.5cm, 9cm, 5.5cm}, clip]{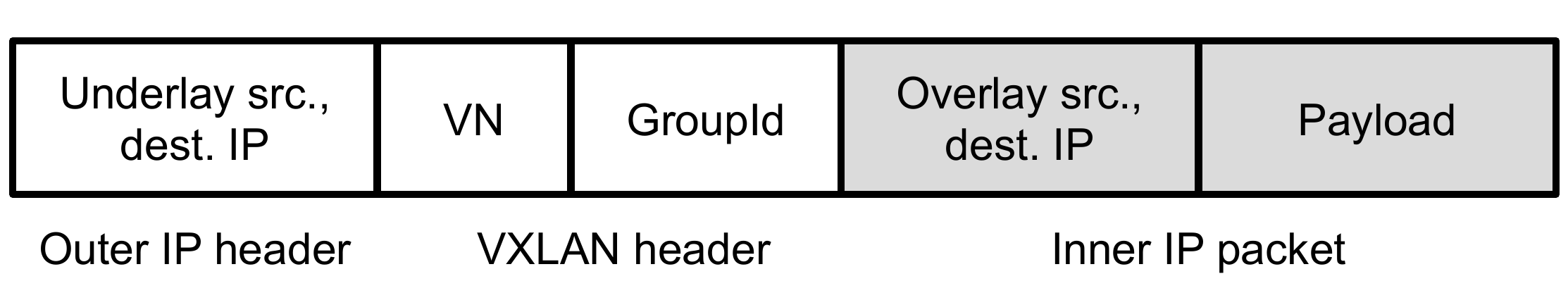}
\caption{Packet Format }
\label{fig:format}
\end{figure}

The data plane encapsulation leverages  VXLAN with Group Identifier extension \cite{smith-vxlan-group-policy-05}. We chose this encapsulation over the native LISP data plane because of the need to encapsulate both L2 and L3 payloads (LISP supports L3 only). Additionally, in  this VXLAN variant  we can add the source GroupId  in the group  field (fig. \ref{fig:format}). Finally, the underlay is a network with plain IP connectivity that routes encapsulated packets between edge routers. The underlay routing is provided by either OSPF or IS-IS, we leverage MACsec \cite{ieeeMACsec2018} for packet integrity protection and confidentiality, and ECMP  for redundancy \cite{ecmpRFC}. The rest of this section describes how the network plugs a device for the first time (onboarding) and the standard packet flow.

\subsubsection{Host Onboarding}\label{sec:sec:onboarding}
\begin{figure}[!tp]
\centering
\includegraphics[ height= 4.5cm, keepaspectratio , trim={0cm, 5cm, 12cm, 0cm}, clip]{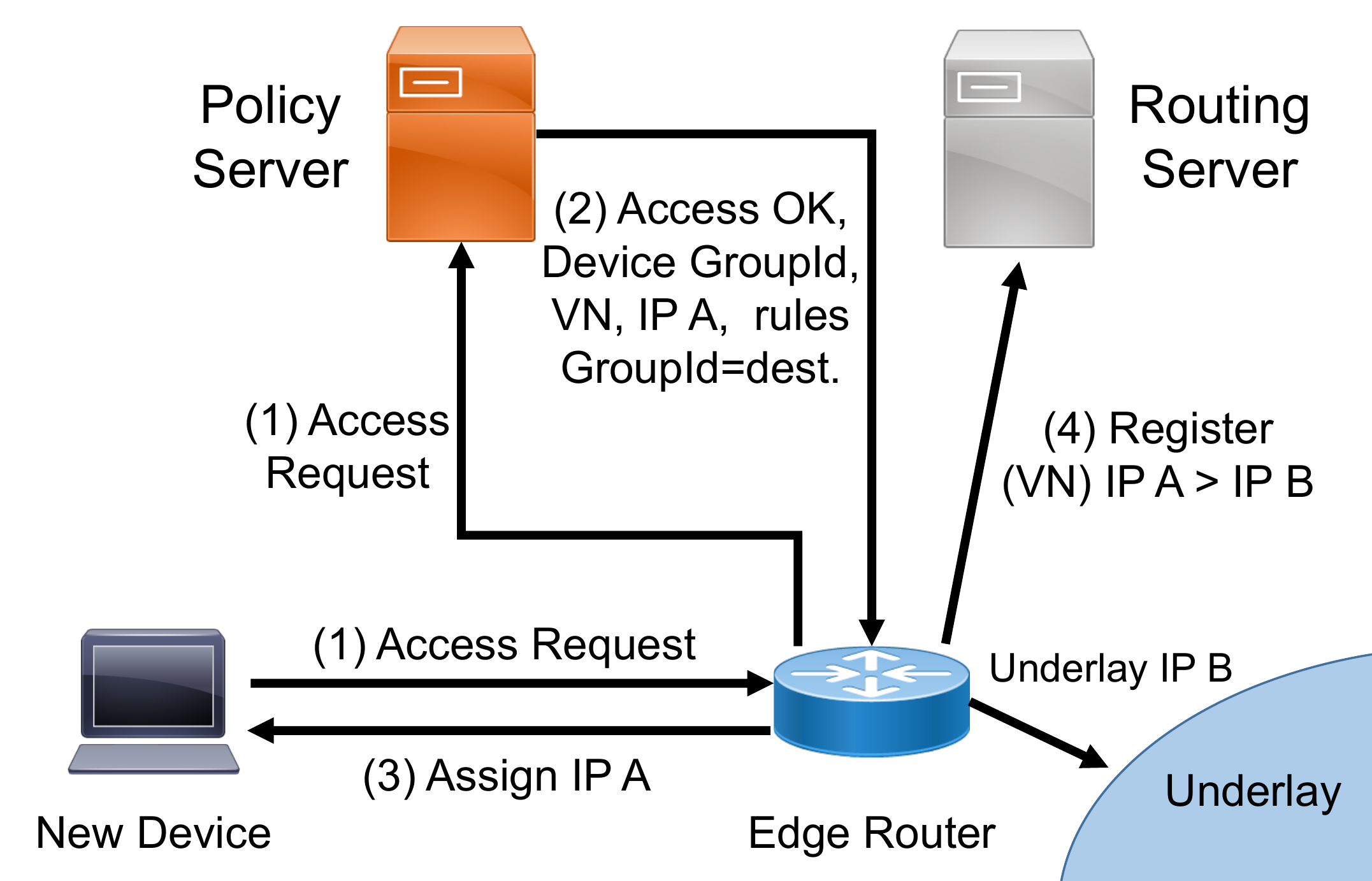}
\caption{Host Onboarding Process}
\label{fig:onboarding}
\end{figure}

When an endpoint connects to the overlay, the edge router detects it and starts the authentication process with the policy server (step 1 in fig. \ref{fig:onboarding}).  After a successful authentication, the edge router downloads the endpoint's VN, GroupId, and associated connectivity rules  from the policy server (step 2). Specifically, it downloads the rules where the endpoint's group is the destination (c.f. sec. \ref{sec:eval:policies}), stores locally the GroupId value, and associates it to the switch port where the endpoint is connected. Then it can assign an overlay IP address to the endpoint (step 4), obtained from a DHCP server (step 3). Finally, the edge router stores the location of the endpoint in the control plane, i.e. update the (VN + overlay IP, underlay IP) pair in the database (step 4). 

\subsubsection{Packet Flow}
\begin{figure*}[!tp]
\centering
\includegraphics[ width= \textwidth, keepaspectratio , trim={0cm 0cm 0cm 10cm}, clip]{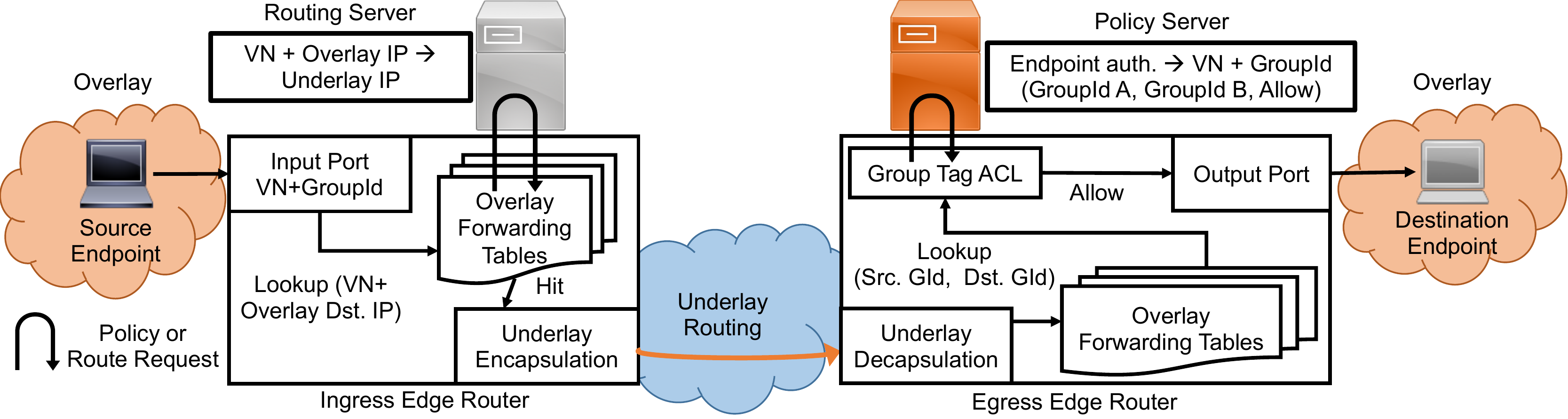}
\caption{Ingress and egress pipelines }
\label{fig:pipeline}
\end{figure*}

On ingress, packets from endpoints are assigned their corresponding GroupId and VN (fig. \ref{fig:pipeline}, ingress edge router). The router knows these values from the onboarding process. Then, it does a VN + overlay destination IP lookup in the VRF table for that VN. If there is no match, it will query the control plane database. This query returns the underlay IP address of the destination overlay IP. Finally, the packet is encapsulated towards the corresponding edge router, carrying both VN and GroupId.

On egress, the destination edge router decapsulates the packet and injects it into a two-stage pipeline (fig. \ref{fig:pipeline}, egress edge router). First, it performs a VN + overlay destination IP lookup in the local VRF table corresponding to the VN in the packet. This query returns the output port and the associated destination GroupId. Each entry in the VRF has its associated GroupId, that is stored during the onboarding process. After authentication, the edge router creates an \texttt{(Overlay IP, GroupId tag)} association in the VRF table.

The second stage is an exact match lookup in an group-based ACL of (source GroupId, destination GroupId). This ACL enforces the aforementioned group rules. Finally, the router forwards the packet to the destination overlay IP address. We perform the policy enforcement on egress due to increased scalability (sec. \ref{sec:eval:policies}).

\subsection{Mobility Support}\label{sec:sec:mobility}
\begin{figure}[!tp]
\centering
\includegraphics[width=0.7\columnwidth, trim={2cm, 1cm, 15.5cm, 6cm}, clip]{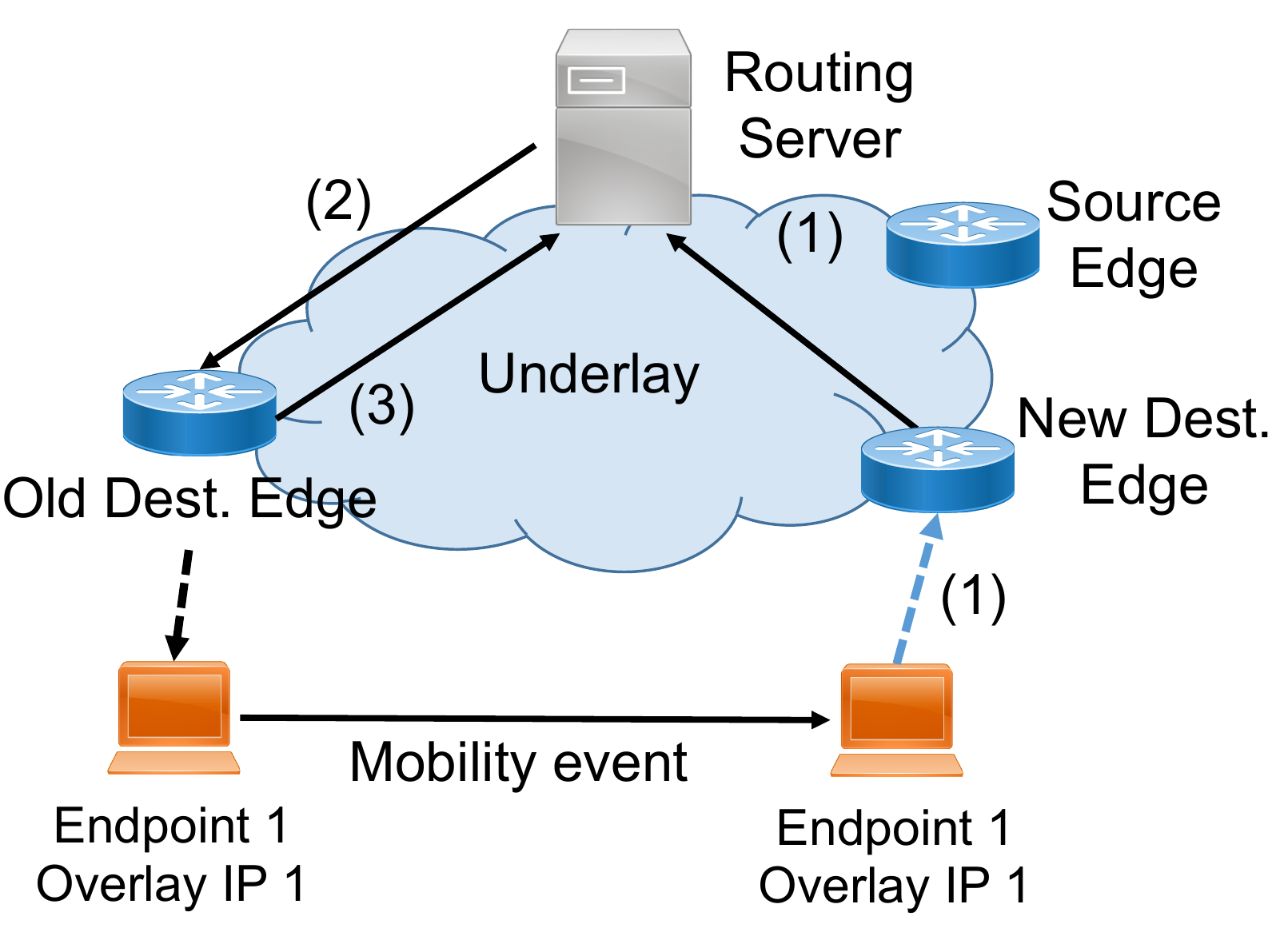}
\caption{Endpoint mobility}
\label{fig:mobiliy}
\end{figure}

When an endpoint roams and attaches to a new edge router, the latter triggers the authentication process again, and registers the new location (Fig. \ref{fig:mobiliy}, Step 1). After this registration, the control plane sends a message to the previous edge router instructing it to (i) pull the new location data (2,3), and (ii) forward all traffic for that particular endpoint to the new edge router. Hence, handover signaling is linear with the number of roaming endpoints, as opposed to proactive protocols, in which it also depends on the number of routers (sec. \ref{sec:sec:massiveMobility}).

\begin{figure}[!tp]
\centering
\includegraphics[width=0.7\columnwidth, trim={2.5cm, 3cm, 14.5cm, 5cm}, clip]{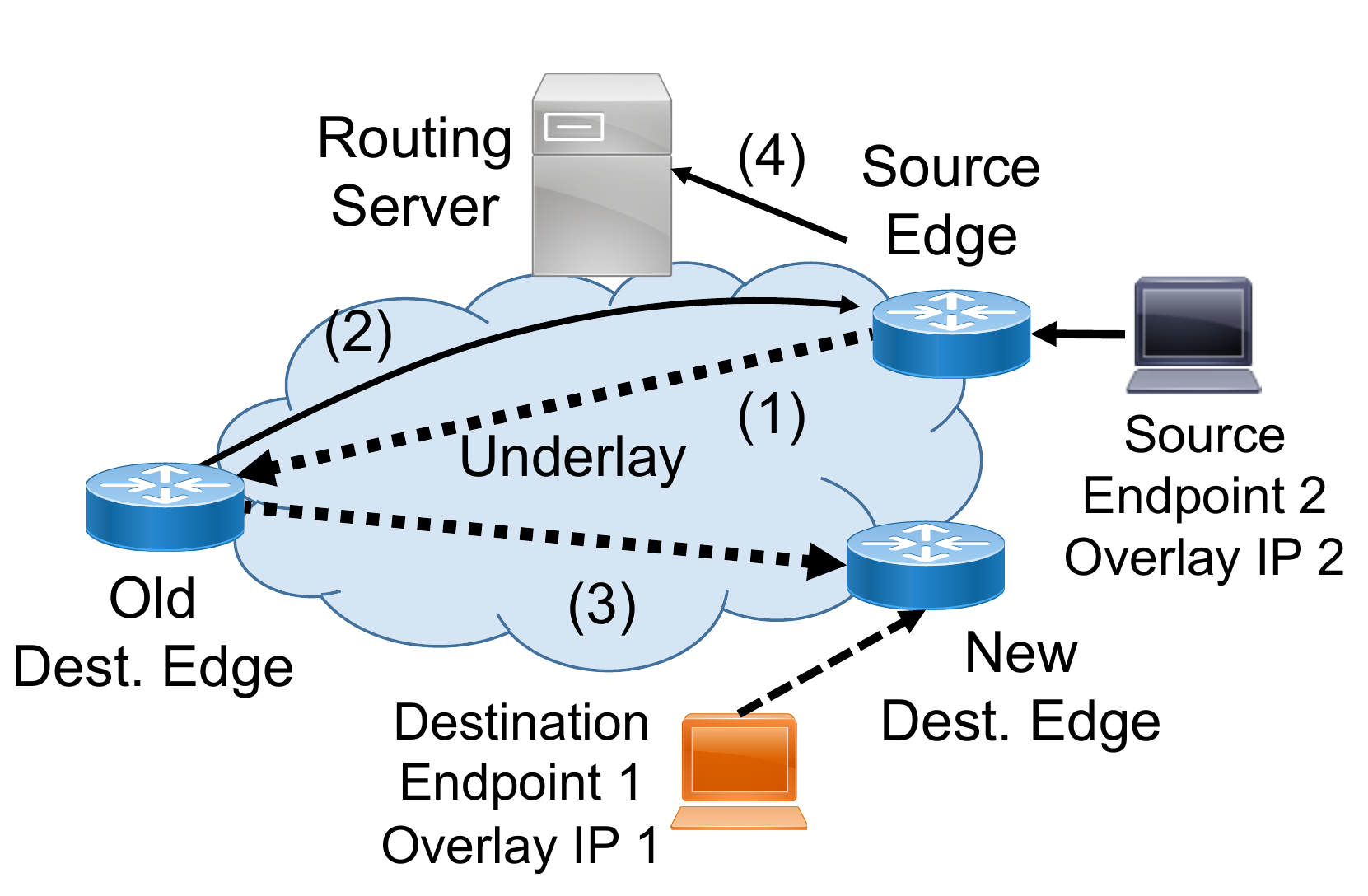}
\caption{Updating stale entries via data plane messages.}
\label{fig:dataplane_trigger}
\end{figure}

Additionally, we apply the event-driven approach to update edge routers storing stale entries. To this purpose, we use a specific data-triggered control plane message (Fig.  \ref{fig:dataplane_trigger}): when the old edge router receives traffic for the roaming endpoint (1) it sends a control message to the source (2), instructing it to retrieve the new location (4). At the same time, the old destination router forwards this traffic to the new destination router (3). 

Regarding signaling scalability, this method depends on traffic patterns: if the roaming endpoint is very popular, we will have to update a significant portion of edge routers. On the contrary, endpoints that receive traffic from few sources, require less signaling. The advantage of this technique is that it is triggered by traffic, in other words, the control plane doesn't need to update \emph{all}  edge routers that have the stale location, but only those that \emph{require} it. Finally, these control plane messages will be staggered over time as traffic from different senders will arrive at different times, thus spreading the signaling load in time.

\vspace{-2.5mm}

\subsection{Support for L2 services}\label{sec:sec:l2} In enterprise scenarios, it is common that some devices or users require L2 connectivity. Common use-cases are some forms of load-balancing, certain IoT devices, and basic services such as DHCP or service discovery\footnote{A significant amount of applications rely on broadcast domains, e.g. Apple Bonjour}.

In order to support such services, but avoid sending broadcast traffic to the entire network, our implementation leverages four elements: (i) VLANs (limited to the edge router ports), (ii) indexing endpoints by MAC address in the Routing Server (in addition to IP address), (iii) storing overlay IP to MAC pairs in the routing server, and (iv) deployment of L2 gateways in edge routers.

The combination of these four elements helps us to provide scalable L2 connectivity: first, VLANs limit broadcast domains; second, MAC address indexing locates endpoints of the same VLAN that are in different edge routers. Finally, L2 gateways absorb broadcast traffic and convert it to unicast: for instance, they capture ARP requests and perform a lookup in the routing server to find the MAC associated to the IP in the ARP request. Then they use this MAC to replace the broadcast MAC in the ARP request, creating a unicast L2 message. This message is injected in the L2 pipeline, which will use the MAC-to-underlay IP to encapsulate the request to the intended L2 MAC.

\section{Evaluation}\label{sec:eval}

We have implemented SDA in a commercially available line of routers, leveraging the protocols mentioned previously: LISP \cite{ietf-lisp-rfc6833bis-27}, RADIUS  \cite{rfc2865}, and Scalable-Group Tag eXchange Protocol \cite{sxp} for the control plane, and VXLAN \cite{rfc7348} for the data plane. We have deployed our implementation in two different real-life scenarios: two campus networks with 150 and 450 endpoints, and a large warehouse (partially emulated) with massive mobility serving 16,000 emulated endpoints. Table \ref{tab:deployments} presents a summary of their characteristics. 

We evaluated three key elements of SDA: first, the response of the routing server under stressful conditions, because it is critical in the process of establishing data flows. Second, quantifying the state optimization in the data plane due to the edge-border design. And third, assess the difference between an proactive and a reactive protocol in face of massive mobility events.

\begin{table}[!tp]
\small
\caption{Deployments used for evaluation}
\begin{center}
\begin{tabular}{|c|c|c|c|c|}
\hline
\multirow{2}{*}{\textbf{Deployment}} & \textbf{\# Border }    & \textbf{\# Edge } &  \textbf{Number of } \\
                    & \textbf{ Routers}    & \textbf{ Routers} &  \textbf{ endpoints} \\
\hline
Building A & 1 & 7& 150\\
\hline
Building B &  2 & 6 &  450\\
\hline
\multirow{2}{*}{Warehouse} & \multirow{2}{*}{2} &\multirow{2}{*}{200}  & 16,000 \\
 &  &  &  (emulated)\\

\hline
\end{tabular}
\label{tab:deployments}
\end{center}
\end{table}

\subsection{Routing Server Scalability}

The routing server is a critical part of the design, because it allows establishing communication between any pair of endpoints. Because of this, we evaluated its performance depending on the number of routes and queries per second. 

\begin{figure*}[!ht]%
   \begin{subfigure}[b]{0.33\textwidth}
        \includegraphics[width=\textwidth]{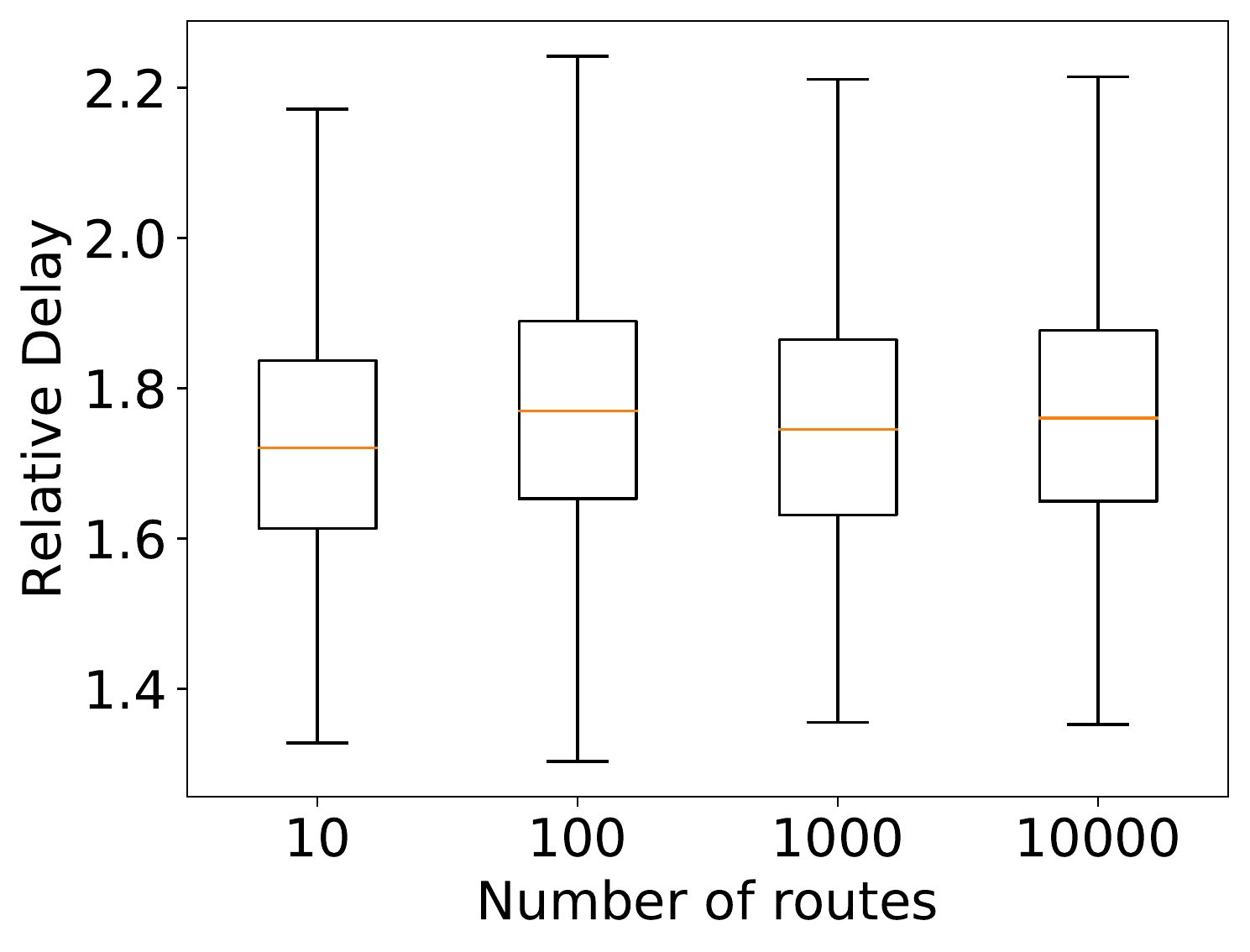}
        \caption{Boxplot (95\%) of the delay of 10k route requests for four different number of configured routes. Relative values to a routing server with 1 route.}
        \label{fig:mrequest_vs_num_elem}
    \end{subfigure}
    \hfill 
    \begin{subfigure}[b]{0.32\textwidth}
        \includegraphics[width=\linewidth]{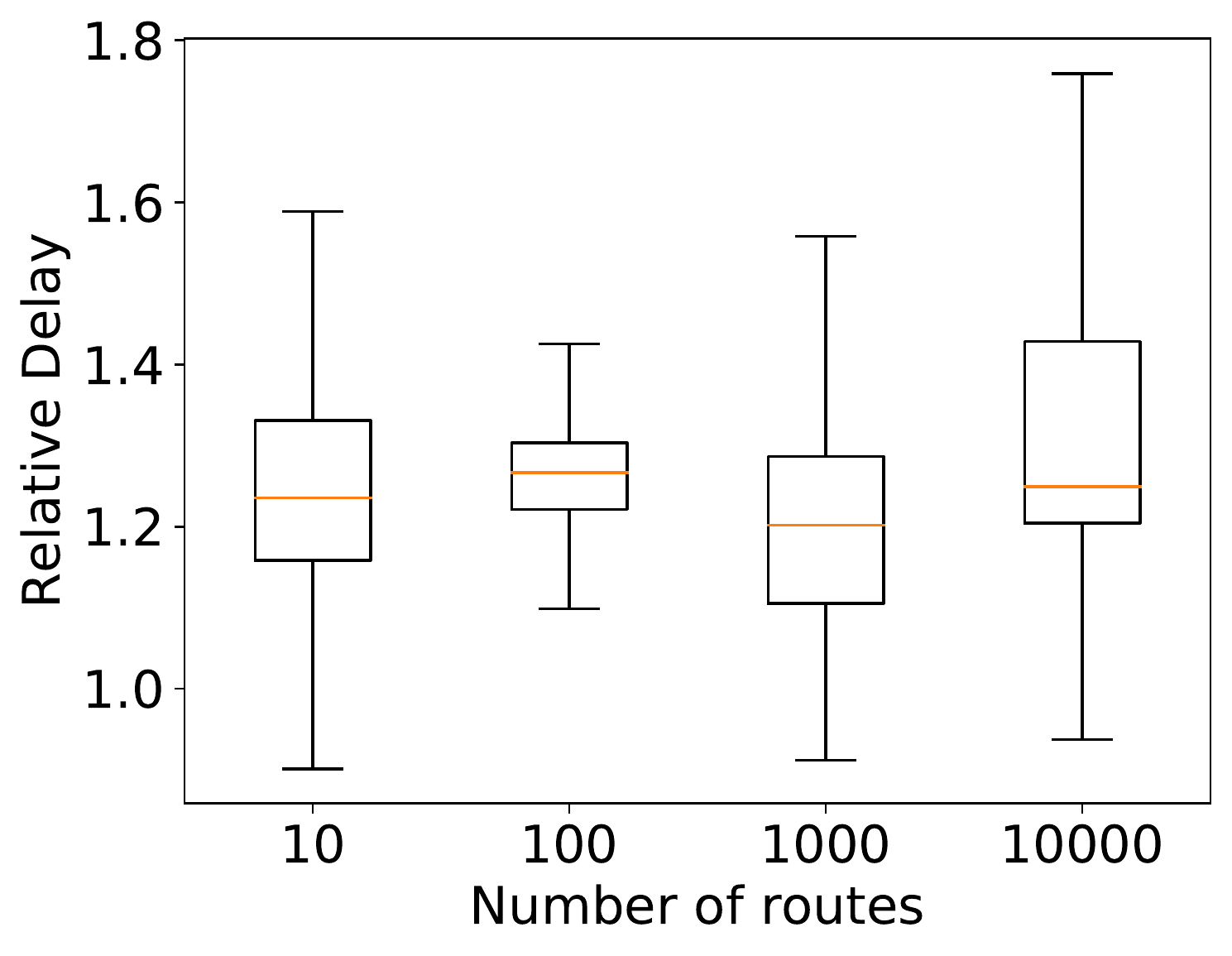}
        \caption{Boxplot (95\%) of the delay of 10k route updates for four different number of configured routes. Relative values to a routing server with 1 route.}
    \label{fig:mregister_vs_num_elem}
    \end{subfigure}
    \hfill
    \begin{subfigure}[b]{0.33\textwidth}
        \includegraphics[width=\textwidth]{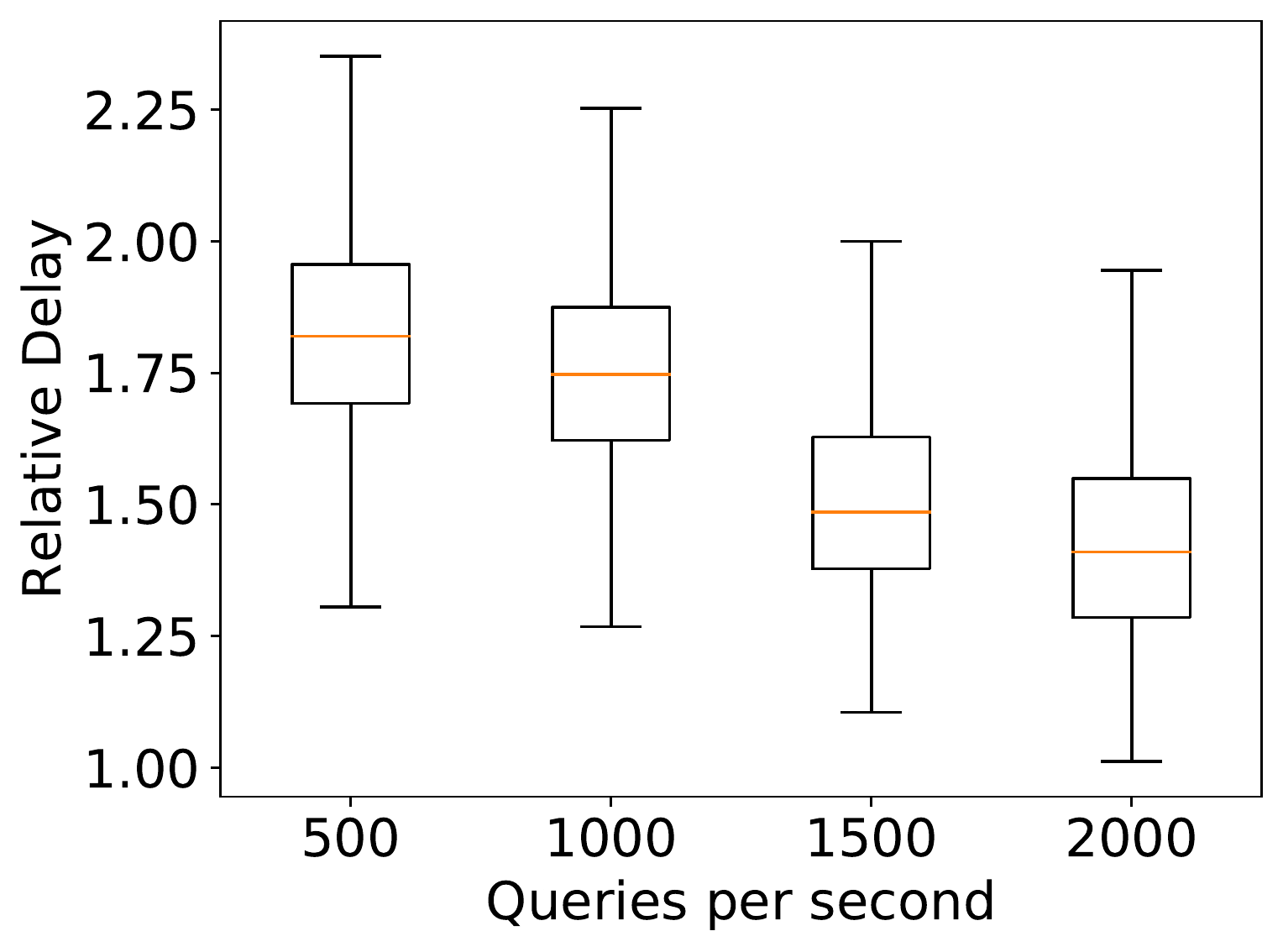}
        \caption{Boxplot (95\%) of the delay of route requests for four different number of queries per second. Values are relative to the minimum of all. }
\label{fig:mreq_num_queries}
    \end{subfigure}%
    \caption{Routing Server Performance Evaluation.}
\end{figure*}

To this end, we setup a routing server implemented in a commercial virtual router with 8 GB RAM, 8 vCPU, on top of a virtualization platform with an 8-core 2.1 Ghz processor. We  measured the delay to answer route requests and route updates with a script running in a local machine that sent 800 queries per second. We repeated the experiment for different number of routes configured in the routing server. Each query requested or updated a different route, in order to avoid optimizations due to intermediate caches. We consider the network delay negligible. Figures \ref{fig:mrequest_vs_num_elem} and \ref{fig:mregister_vs_num_elem} present boxplots of the time required to answer an IPv4 route request and a route update, respectively, for four different number of configured routes in the routing server. The values are relative to the minimum delay of a routing server with only one route. 

We can see that the delay is not dependent on the number of routes. Since this architecture is designed to store network state hierarchically, it makes it easy to implement the routing server with a Patricia Trie. The delay of this data structure depends on the number of bits of the keys, not the number of elements \cite{morrison68patricia}. Based on the data collected for this particular test, an equivalent deployment using a similar setup should scale to at least 3k endpoints without noticeable performance degradation. Each endpoint requires registering 3 routes (IPv4, IPv6 and MAC addresses), then 10k / 3 = $\sim$3k. 

We chose to send 800 queries/s to the server since it is the peak requirement in the massive mobility scenario (sec. \ref{sec:sec:massiveMobility}). We consider this a highly loaded server, however, in case we needed to increase such figure, the architecture scales horizontally and can deploy more routing servers. Then, we load balance across edge routers by grouping them and pointing each group to a different routing server for the route requests, and perform route updates on all servers.

Using the same routing server as before, we repeated the previous experiment but for different number of queries/s. Figure \ref{fig:mreq_num_queries} presents boxplots of the delay to answer IPv4 route requests for four different number of queries per second to the routing server. The values are relative to the minimum delay of all samples. Assuming a mobility pattern similar to the warehouse scenario (sec. \ref{sec:sec:massiveMobility}) with 800 moves/s and that each move triggers 2 queries to the routing server, we conclude that the routing server could support this use case (800*2 = 1600 queries/s).

\subsection{State Reduction}
\begin{figure}[!tp]
\centering
\includegraphics[ trim={14.75cm 11cm 2.5cm 2cm}, clip, width=\columnwidth
]{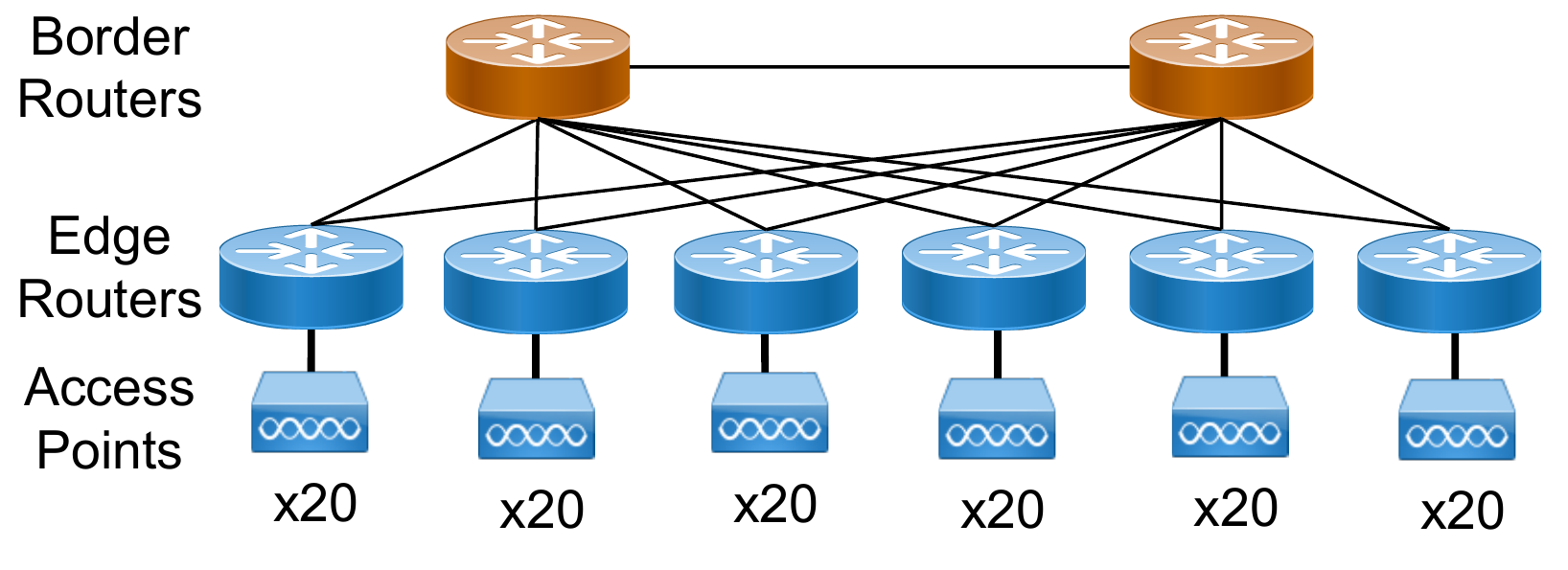}
\caption{Campus Network Topology (underlay routers not shown for clarity)}
\label{fig:campustopo}
\end{figure}

In order to quantify the state reduction due to the reactive protocol, we counted the number of overlay-to-underlay IPv4 mappings in the FIB of the edge and border routers. We setup a VM that collected this data hourly from the router CLI. The routers were in two separate buildings (A and B), with three floors each, and providing network connectivity to between 200 and 500 users. Table \ref{tab:campus} provides additional details about each deployment, and Fig. \ref{fig:campustopo} shows the network topology. The control plane of border routers has a 4-core 2.4GHz, x86 CPU with 16 GB RAM, and edge routers a 4-core 1.8 GHZ, x86 CPU with 8 GB RAM. Both of them use a custom ASIC for the data plane. The border-to-edge links are 10 Gbps and the edge-to-AP 1 Gbps.

\begin{table}[!tp]
\small
\caption{Details of campus deployments}
\begin{center}
\begin{tabular}{|c|c|c|}
\hline
            & \textbf{Bldg. A}   & \textbf{Bldg. B} \\
\hline
\textbf{Border Routers} &   1   & 2 \\
\hline
\textbf{Edge Routers}    &  7   &  6  \\
\hline
\textbf{Floors}          &  3   &  3  \\
\hline
\textbf{AP per floor}     &  40  &  40  \\
\hline
\textbf{Total AP}         &  120  &  120  \\
\hline
\textbf{AP per edge}     &    $\sim$20 &  20 \\
\hline
\end{tabular}
\label{tab:campus}
\end{center}
\end{table}

\begin{figure*}[!ht]%
   \begin{subfigure}[b]{0.33\textwidth}
        \includegraphics[width=\textwidth]{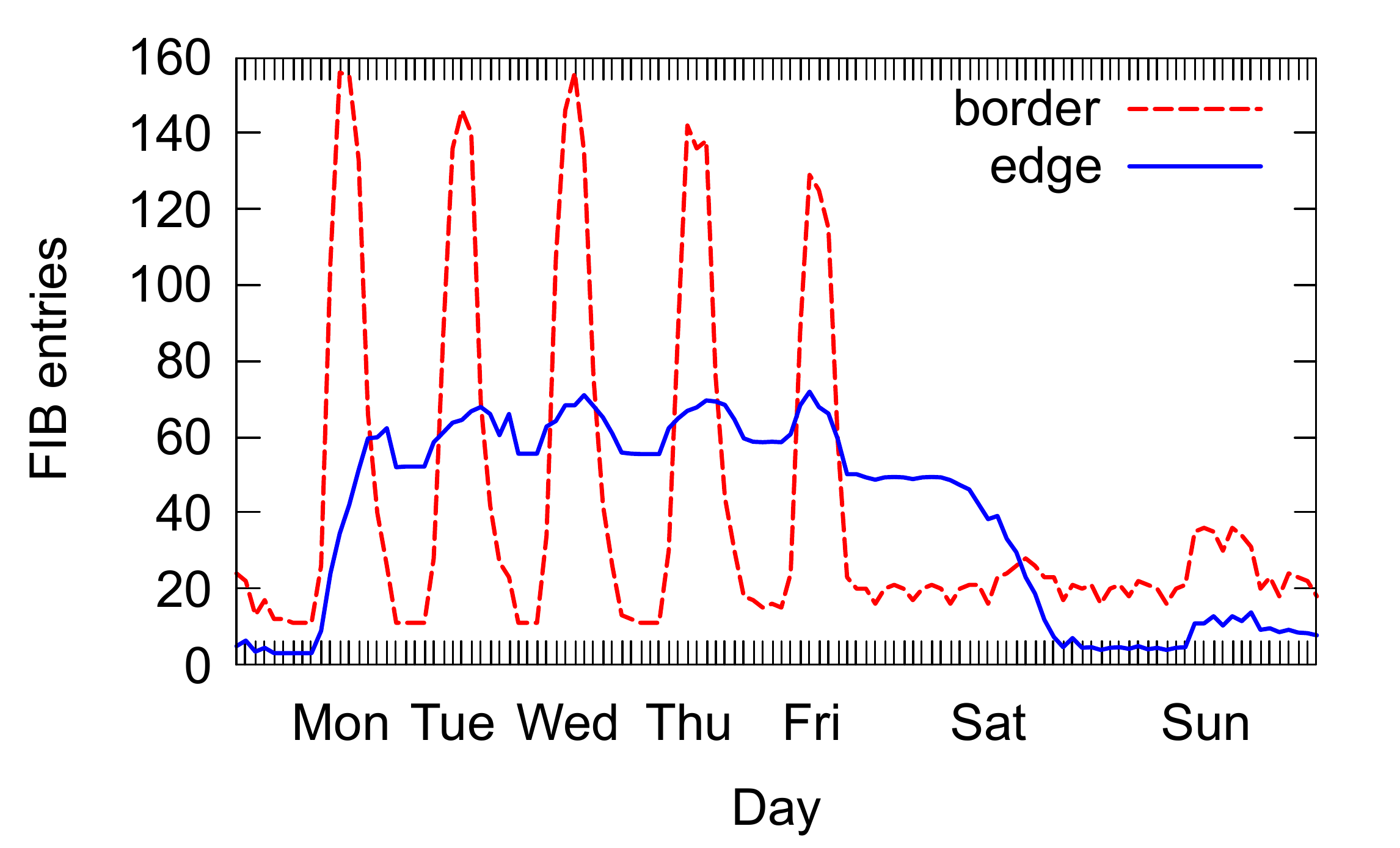}
    \end{subfigure}
    \hfill 
    \begin{subfigure}[b]{0.32\textwidth}
        \includegraphics[width=\linewidth]{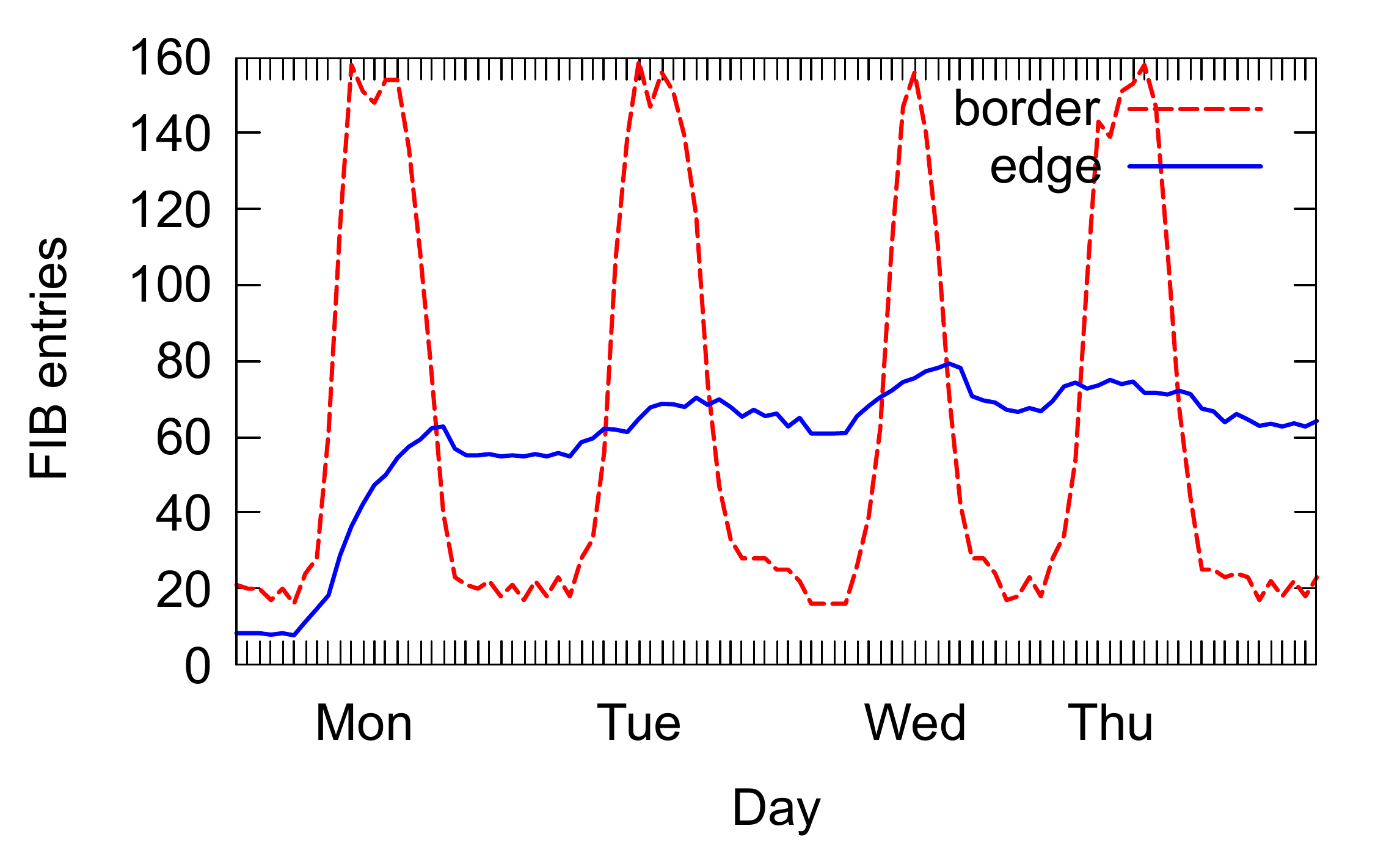}
    \end{subfigure}
    \hfill
    \begin{subfigure}[b]{0.33\textwidth}
        \includegraphics[width=\textwidth]{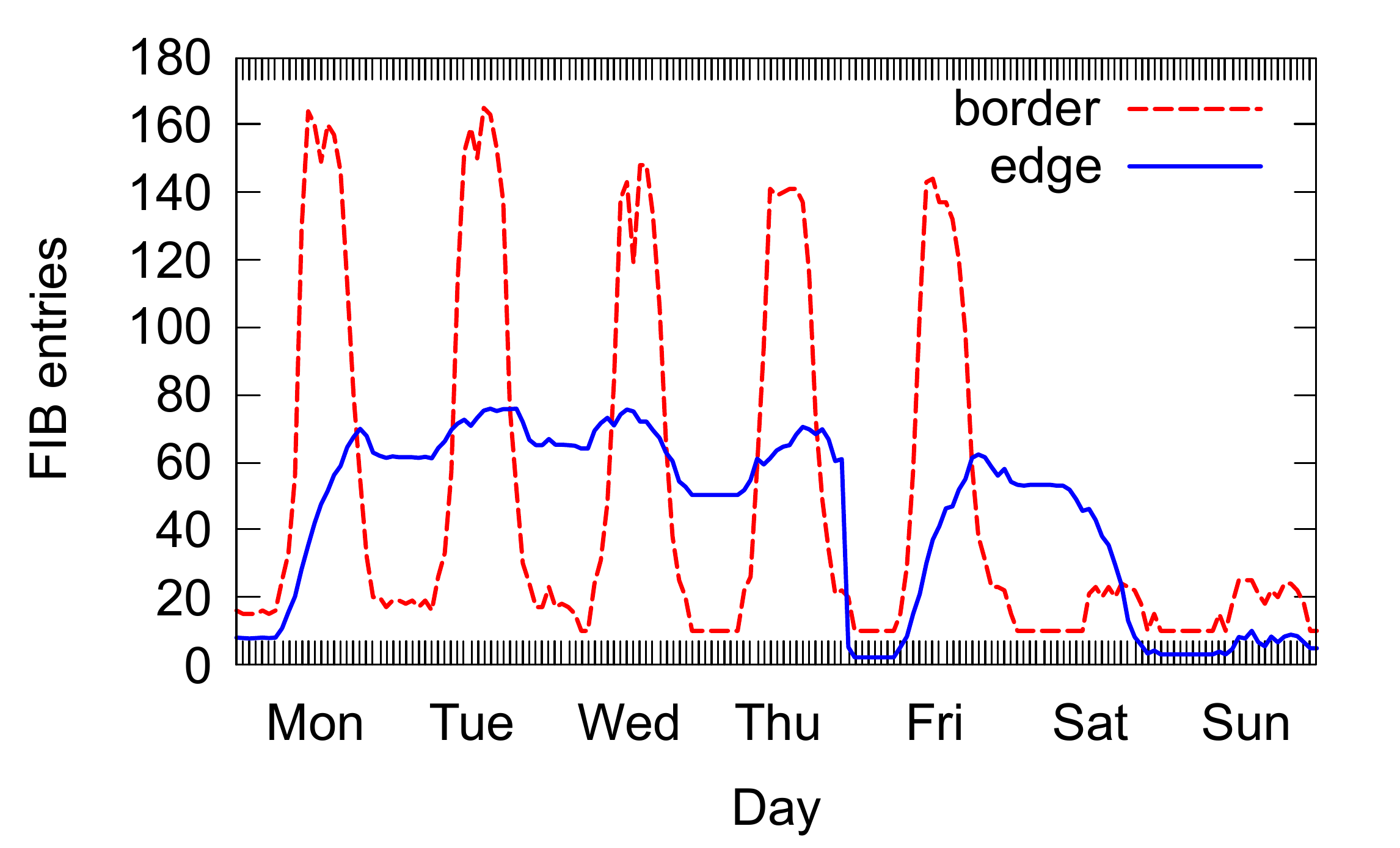}
    \end{subfigure}\\
    
   \begin{subfigure}[b]{0.33\textwidth}
        \includegraphics[width=\textwidth]{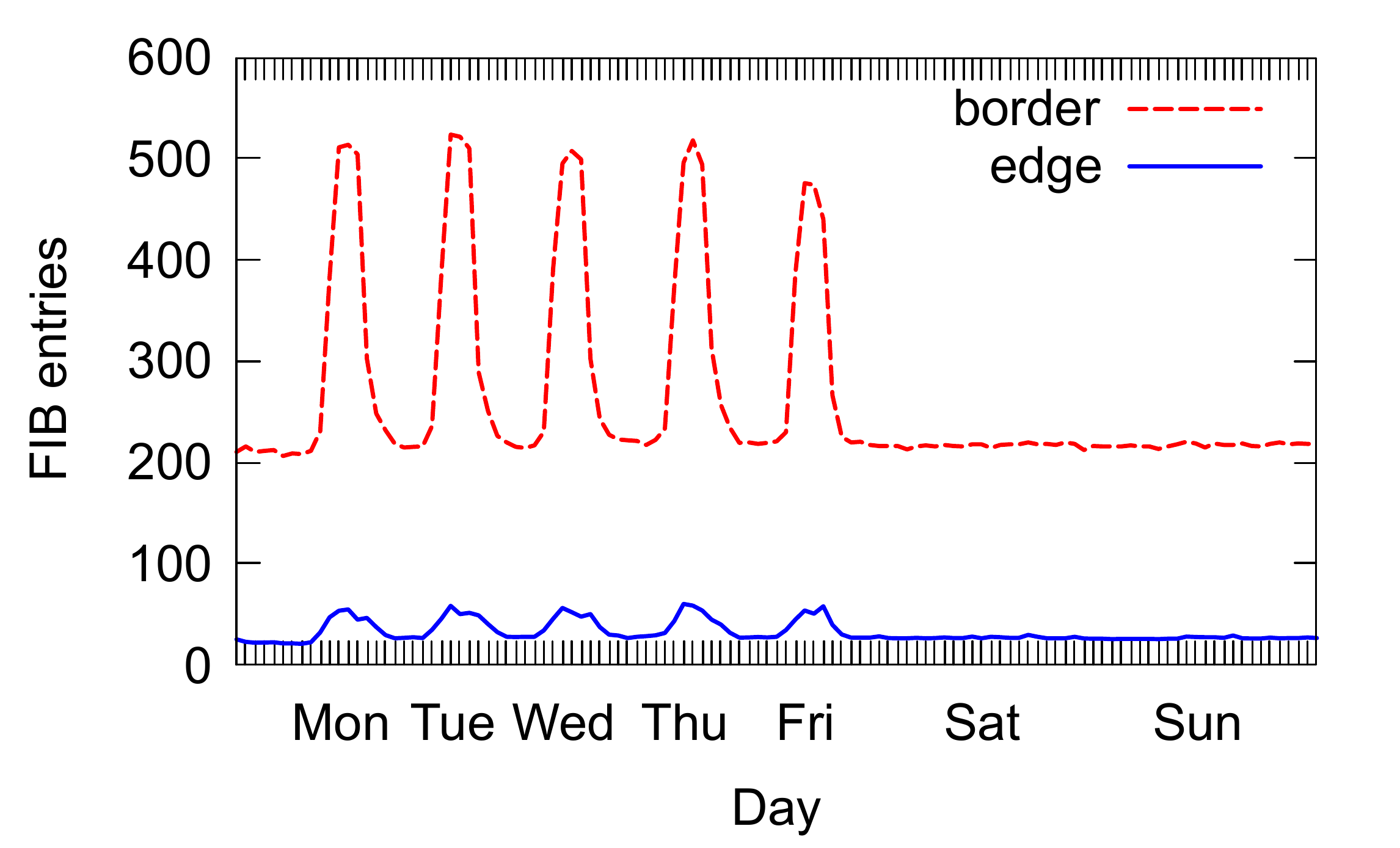}
    \end{subfigure}
    \hfill 
    \begin{subfigure}[b]{0.32\textwidth}
        \includegraphics[width=\linewidth]{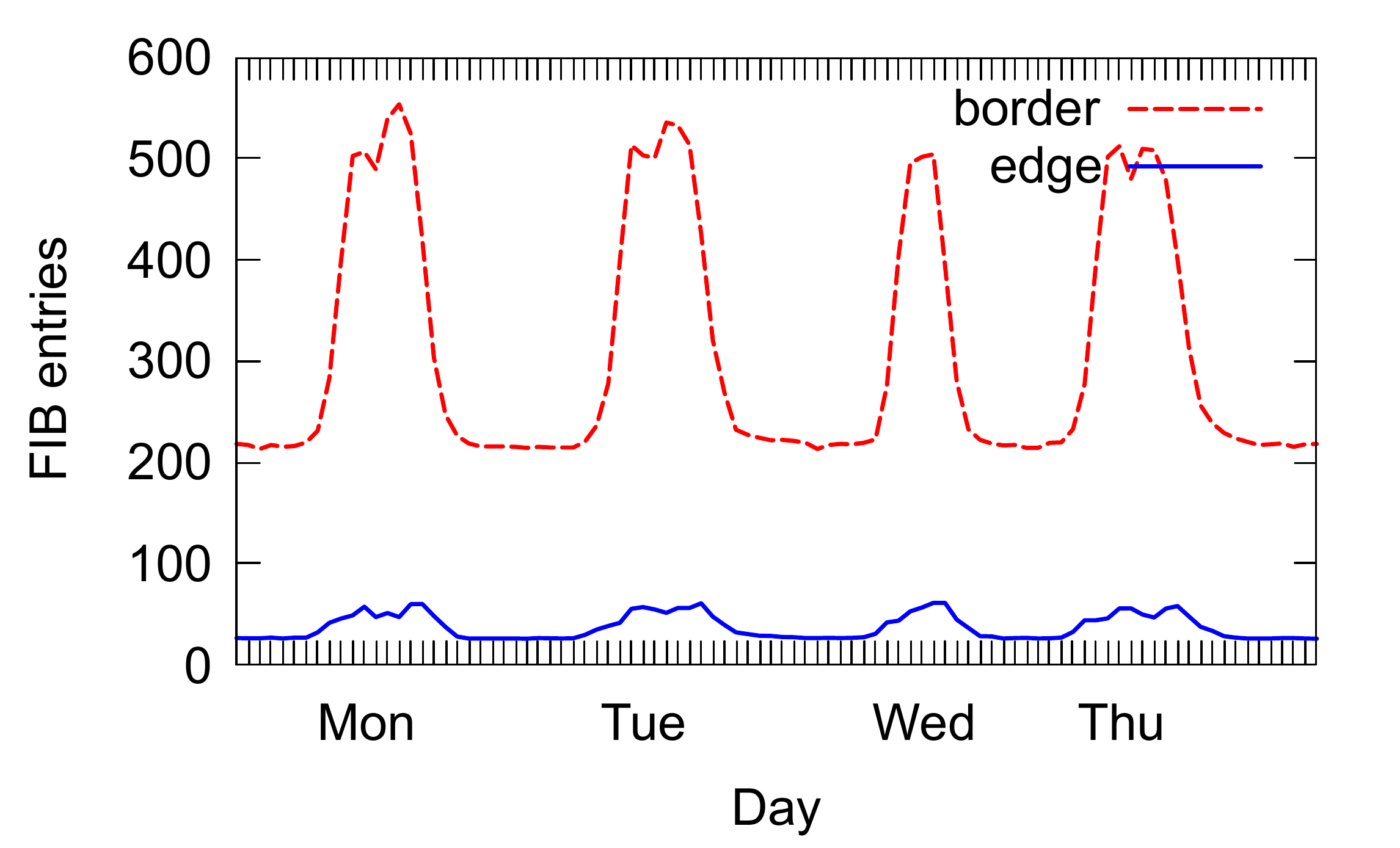}
    \end{subfigure}
    \hfill
    \begin{subfigure}[b]{0.33\textwidth}
        \includegraphics[width=\textwidth]{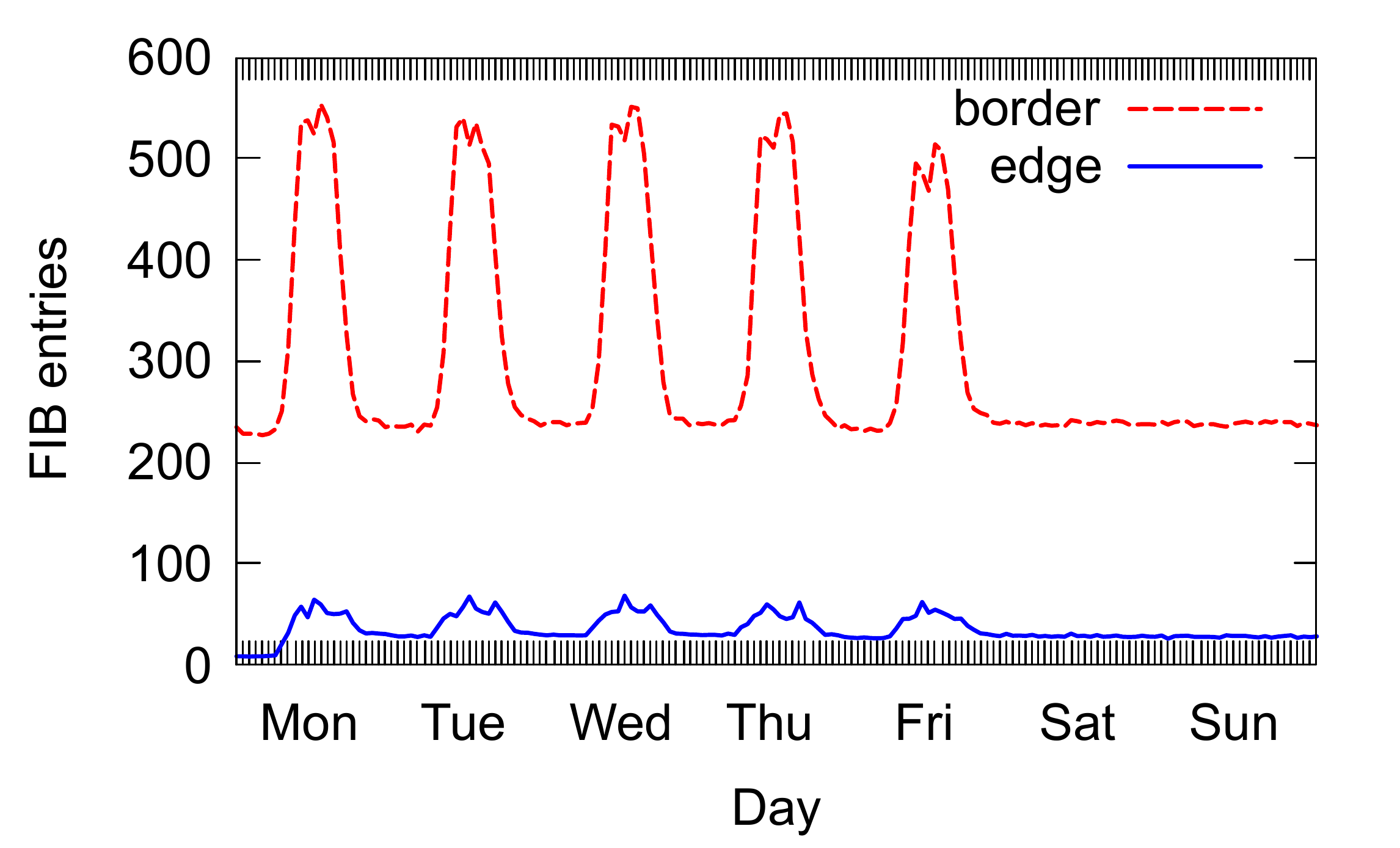}
    \end{subfigure}%
    \caption{Number of FIB entries in border vs. edge router for three different weeks. Top row corresponds to building A, and building to B.}
    \label{fig:fig_fib}
\end{figure*}

Figure \ref{fig:fig_fib} shows the average number of FIB entries for the border and edge routers, for both buildings, and for three different weeks. We can see that, on average edge routers store less FIB entries than border routers: in building A edge routers carry only about 30\% of the FIB entries on the border routers, while in building B this figure is as low as 6\%. Table \ref{tab:fib_averages} presents also averages of FIB entries for border and edge routers for a period of 5 weeks. We can also see reductions in FIB entries, even if we calculate separate averages for day and nighttime. We can conclude that the reactive approach helps in optimizing data plane state, while the border router absorbs the convergence delay of the route resolution. 

It is also worth mentioning that the usage pattern of the FIB table on border routers shows a common daily and weekly pattern: during daytime in working week days  routers host more routes than during nighttime and over  weekends. This is due to the fact that the border router is always up to date with the information in the routing server regarding the endpoints in the deployment.  Thus, the number of entries in the border router follows closely the presence of authenticated users in the network (i.e. users in the office). In contrast, edge routers cache routes learned on demand and may retain them during longer periods, even when users have left the office. We can clearly see this in building A (Fig. \ref{fig:fig_fib}, top row), where edge routers seem to keep their routes for most of the time between workdays, but eventually clear they caches during weekend. On the other hand, this effect is less noticeable in building B, where edge routers follow the daytime/nighttime routine more closely. The reason behind this could be nighttime traffic patterns: when some endpoints leave the office at night, the remaining ones may initiate communication with those that left, that will trigger a route resolution with a negative result, and thereby deleting that FIB entry.

Finally, another relevant aspect of building B is a substantial amount of end-hosts that are permanently connected to the network and do not follow the day/night routine. Examples of such end-hosts are desktops and IoT devices (VoIP phones, cameras) that do not necessarily move with the users.

\begin{table}[!tp]
\small
\caption{Average number of FIB entries for a 5 week period, for day work hours (9 am to 19 pm), and nighttime. }
\begin{center}
\begin{tabular}{|c|c|c|c|}
\hline
\multirow{2}{*}{\textbf{Router}} & \multirow{2}{*}{\textbf{Period}}   & \multicolumn{2}{|c|}{\textbf{Building}}  \\
\cline{3-4}
&  & \textbf{A}  & \textbf{B} \\
\hline
\multirow{3}{*}{Border}  & All & 50 & 291 \\
\cline{2-4}
& Day & 85 & 362 \\
\cline{2-4}
& Night & 19 & 227 \\
\hline
\multirow{3}{*}{Edge} & All & 42&  34 \\
\cline{2-4}
& Day & 47 & 42\\
\cline{2-4}
& Night & 38 & 27\\
\hline
\multicolumn{2}{|c|}{\textbf{Decrease (All)}}    & 16\%  &   88\%  \\
\hline
\end{tabular}
\label{tab:fib_averages}
\end{center}
\end{table}

\subsection{Massive Mobility Events}
\label{sec:sec:massiveMobility}

\begin{figure}[!tp]
\centering
\includegraphics[width=\columnwidth, trim={8cm 4.5cm 11cm 4.5cm}, clip]{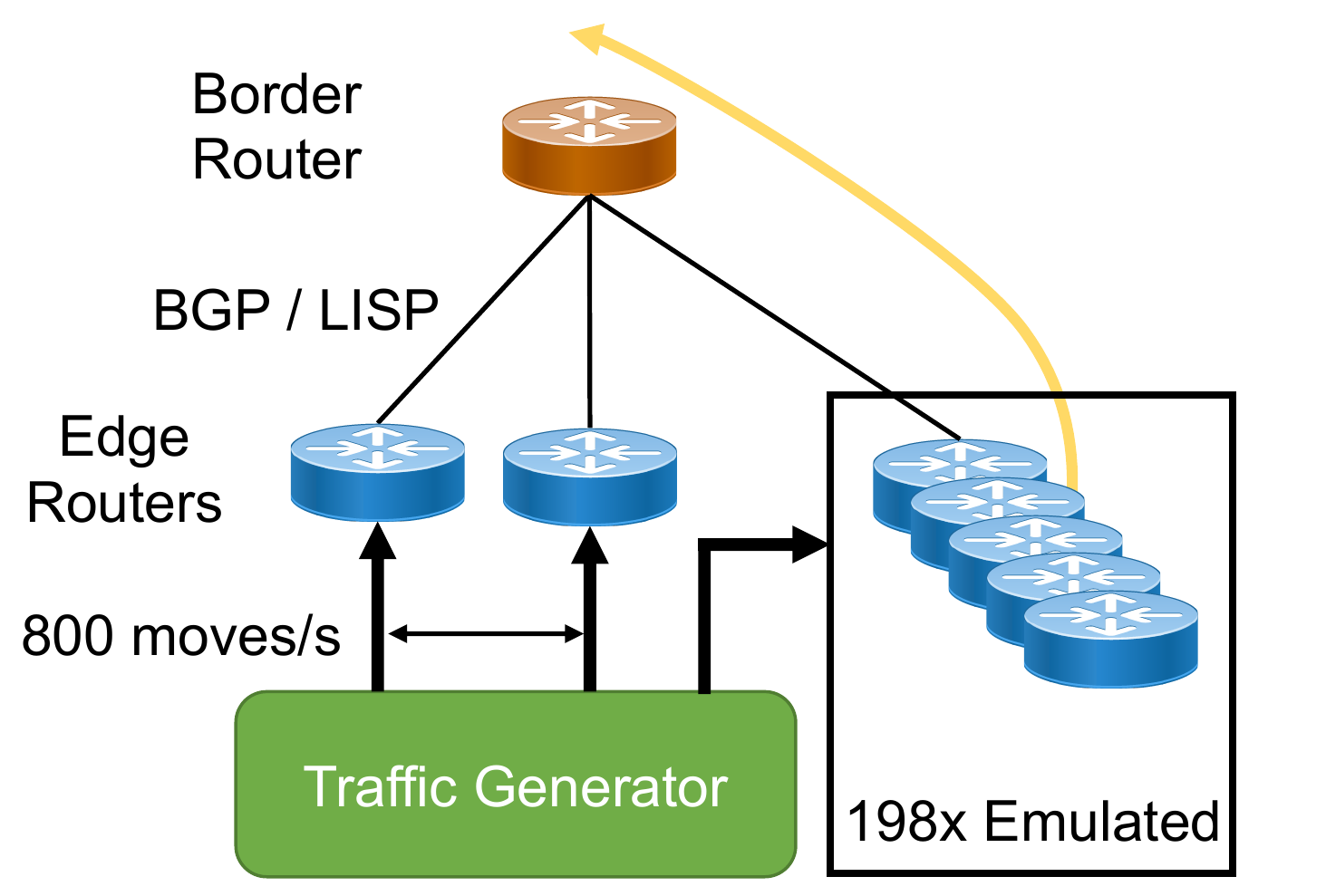}
\caption{Warehouse Network Topology}
\label{fig:mobilitopo}
\end{figure}

In this experiment we focus on the handover delay of large mobility events for a reactive and proactive protocol. We recreated in the lab the specifications of a real life deployment, a large warehouse with hundreds of robots continuously roaming across WiFi access points. Figure \ref{fig:mobilitopo} presents the topology: a commercial border router (4-core 2 GHz CPU, 8 GB RAM for the control plane and custom ASIC in the data plane), two commercial edge routers (4-core 1.8GHz CPU, 8 GB RAM for the control plane and custom ASIC in the data plane) and 198 edge routers emulated with a commercial traffic generator. The border router had an embedded routing server.

We configured the traffic generator to: (i) create unidirectional UDP traffic from the 200 edge routers towards the border router (yellow arrow in fig. \ref{fig:mobilitopo}), (ii) send 1500 bytes packets from 16,000 emulated hosts, and (iii) generate 800 mobility events per second by changing the attachment port of the end hosts between the two physical edge routers. We selected this amount of mobility events because in the real-life scenario, around 5\% of the endpoints change their attachment point every second.

We measured the convergence time as the handover delay, i.e. the time since the emulated host is detached until traffic is restored after it attaches to the new edge router. In order to compare both proactive and on-demand approaches, we measured the handover delay in the same topology, but with two different control plane configurations: BGP as the proactive, and LISP as the reactive. In the BGP case we used a centralized route-reflector in the edge router to distribute route updates.

\begin{figure}[!tp]
\centering
\includegraphics[width=0.9\columnwidth, trim={1cm, 1cm , 2cm, 0cm}]{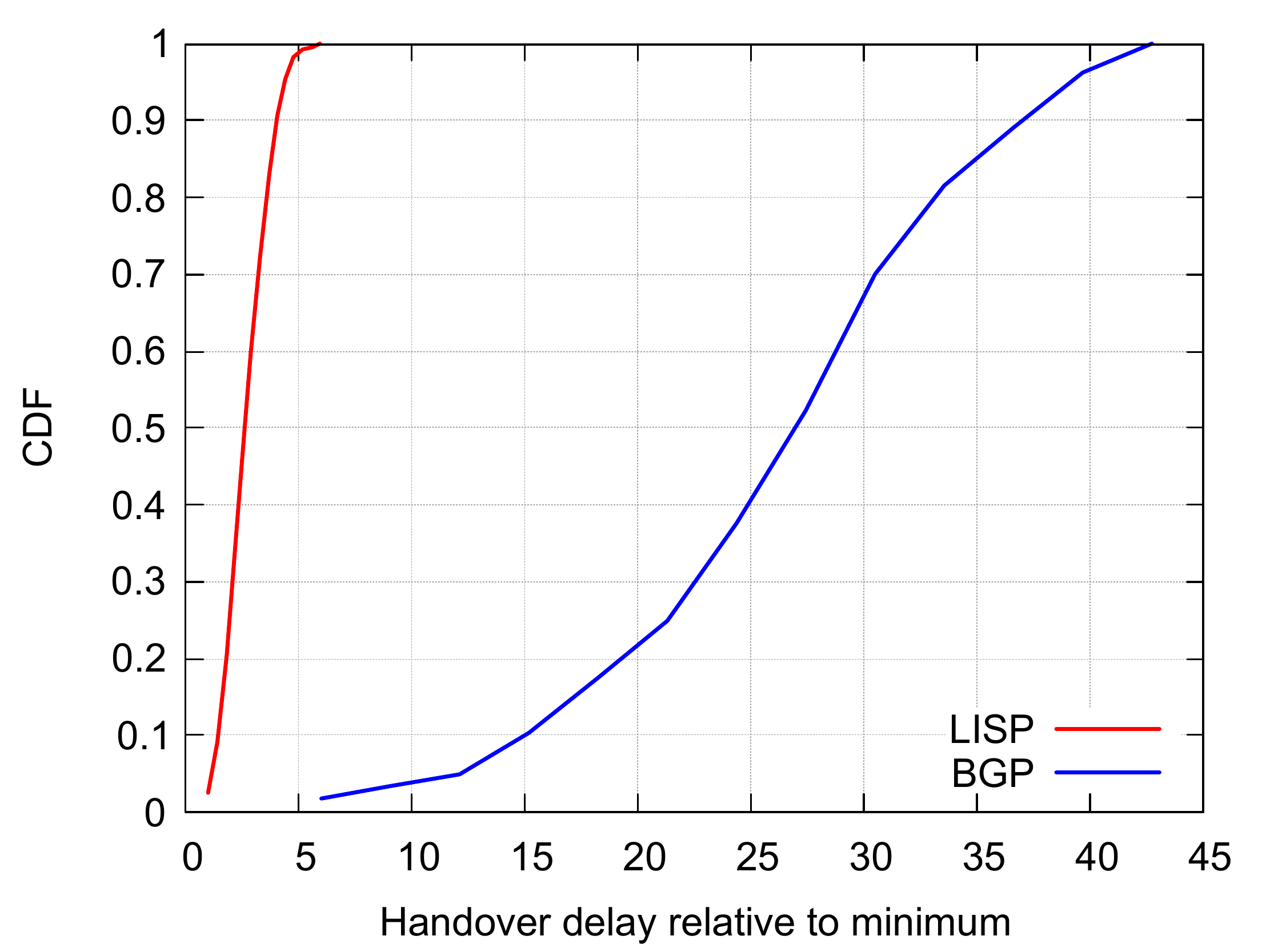}
\caption{Handover Delay for Event-driven (LISP) and Proactive (BGP)  Protocols}
\label{fig:ho_delay}
\end{figure}

Figure \ref{fig:ho_delay} plots the CDF of the handover delay for BGP and LISP. All values are normalized to the minimum convergence time observed during the measurement process.  We can see that the proactive solution takes around 10 times more to converge than the reactive one. The reason is that the proactive approach  replicates the network update to all 200 edge routers, while the on-demand approach follows traffic patterns and only updates edge routers that have active traffic to roaming hosts.

Another important observation is that the variance of the handover delay is consistently higher in the proactive approach than the reactive. This is due to the fact that the reactive architecture selectively updates only routers that are actively sending traffic to the end-host that moved, while the proactive approach updates edge routers randomly, i.e. not by their need for such update. These results show that SDA's reactive approach can be beneficial in stressed environments such as automated warehouses or large gatherings with  highly mobile end-hosts.

\section{Lessons Learnt}
\label{sec:lessons}
In this section we summarize several challenges and our learnings from implementing and  deploying SDA in enterprise networks.

\subsection{Underlay Connectivity Issues}

In order for the overlay to work, there needs to be underlay connectivity. However, it is possible that an edge router fails or that an underlay IP changes, interrupting normal traffic flow. It can be challenging for the overlay to know the state of the underlay without explicit probing. To cope with these situations, edge routers monitor the address announcements of the underlay routing protocol (IS-IS or OSPF) to know about their reachability to underlay IP addresses of the other edge routers. 

This way, when they detect a connectivity outage, they update their local forwarding table deleting such route and falling back to the default route to the border, until a new edge router registers the overlay address in the routing server.

\subsection{Edge Routers Rebooting}

One issue that can happen is a forwarding loop between the edge and border router. Assume the network is forwarding traffic, and an edge router reboots. It will start with an empty FIB for the overlay entries. When it receives traffic for one of its former endpoints, it will use the default route and forward it back to the border router, since it does not know yet its endpoints. The border router will forward this traffic to the rebooted router according to its current information, creating a forwarding loop. 

Although this loop is transient and disappears once the edge router detects its endpoints, we rely on two mechanisms in such situation. First, since the edge router will not announce its underlay IP address through the underlay routing protocol while rebooting, the tracking of the connectivity state of the underlay IP address that the other edge routers perform will remove the routes to the rebooting edge router. Second, the rebooting router will not recognize the incoming traffic, so it will send the data plane message we mentioned in sec. \ref{sec:sec:mobility} to the originating edge router. This will trigger a refresh in the overlay FIB entries of the sending edge router.

\subsection{Selecting the Policy Enforcement Point}\label{sec:eval:policies}
In this section we discuss a bandwidth vs. network state trade-off related to the deployment of group policies. We can save bandwidth by enforcing policies on ingress (because we don't forward traffic that will be dropped on egress), or data plane state by enforcing them on egress. On egress we save data plane state because  we only need policies for the local destination groups of the endpoints that are attached to  a particular edge router, as opposed to ingress, which would need policies for \emph{all} possible destination groups. Taking this into account, we chose to enforce policies on egress to reduce overall state in the data plane.

\begin{figure}[!tp]
\centering
\includegraphics[width=0.9\columnwidth, trim={1cm, 1cm , 1cm, 0cm}]{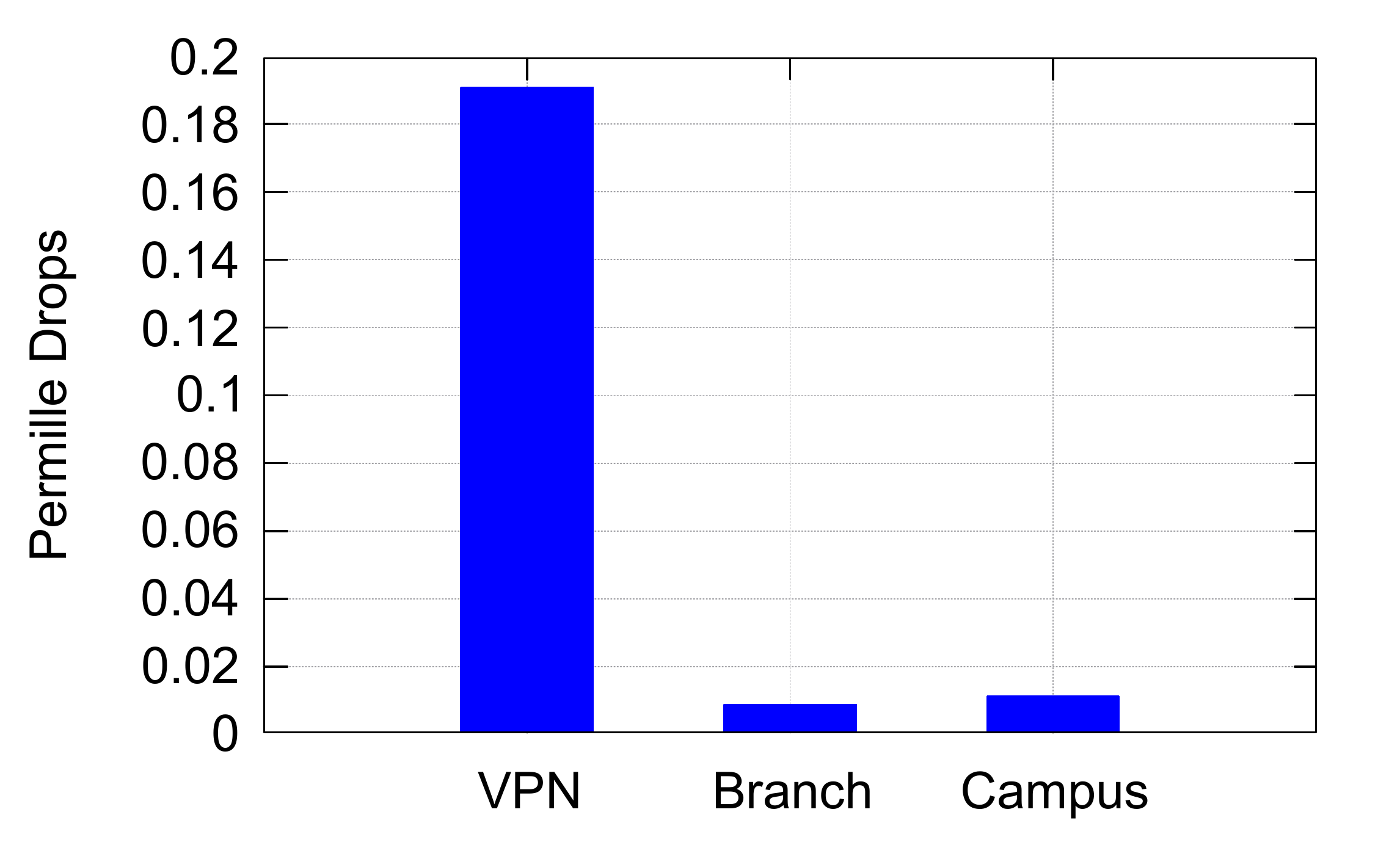}
\caption{Permille hits on drop rules over all hits.}
\label{fig:acl_drop}
\end{figure}

In order to quantify the wasted bandwidth due to enforcing on egress, we analyzed the packet drops due to group-based ACLs in a real-life deployment leveraging egress-based policy enforcement. This deployment is a medium-large enterprise network with a campus and a few branches. For our analysis, we looked at three different devices in this deployment, a branch router, an edge device in the campus, and a VPN gateway. Combined, these three devices were serving around 11,000 endpoints during the period we monitored them. Figure \ref{fig:acl_drop} presents the permille of dropped traffic for the period of 5 days. We can see that in the worst case the drop rate is extremely low: 2 out of each 10k packets. The VPN router has a significantly larger amount of drops than the other routers due to the fact that it receives all the traffic from remote users, which present a different usage pattern from the users in the office.

Since this deployment performs the policy enforcement on egress, we expected a significant percentage of drops. Surprisingly, we discovered that after a new policy is applied, there is a transient period with an increase in drops, but when endpoints (which are usually humans) realize they cannot access this particular destination, they stop requesting it. Hence, the operational experience in the most common enterprise use cases shows  that \emph{enforcing policies on egress does not impact significantly  the amount of wasted bandwidth}.

\begin{figure}[!tp]
\centering
\includegraphics[width=0.8\columnwidth, trim={0cm, 0cm, 14cm, 0.5cm}, clip]{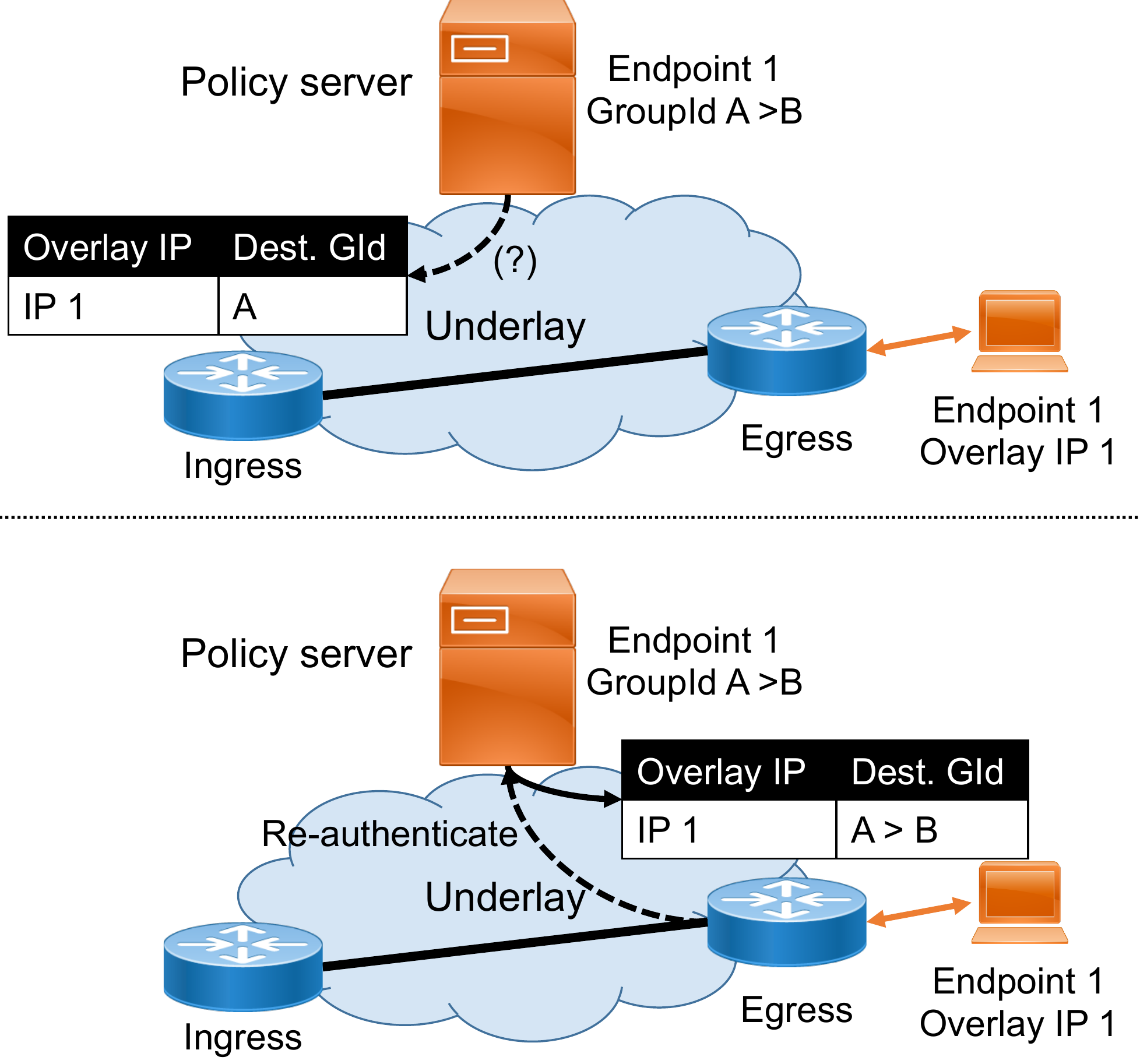}
\caption{Policy Enforcement on Ingress (top) and Egress (bottom)}
\label{fig:updates}
\end{figure}

An additional benefit of enforcing on egress is a simplification of signalling. When a group rule is updated, it is necessary to notify all the affected edge routers. However, this is not easy in our design, because requests to the control plane are only triggered by data plane events. For example, consider that the policy check is implemented on ingress and that the edge router has already learned the GroupId associated to a particular destination (fig. \ref{fig:updates}, top part). Now suppose that the group associated to this destination endpoint is updated, the ingress router has no way to know it. Hence, we need a way to signal this change. 

On the contrary, if the policy is enforced on egress, the (Overlay IP, GroupId) pair in the VRF is automatically updated, because the modification of endpoint data automatically triggers the authentication process again. In other words, on egress the (Overlay IP, GroupId) pair is always up to date because it is linked to the endpoints connected to that edge router. This way, we can avoid implementing an extra signaling mechanism, and the associated complexity.

\subsection{Updating Policies}
In our deployment experience, we found an interesting trade-off when updating policies: it can be more scalable (i.e. less signaling) moving users to different groups rather than directly updating the group-based ACLs. Indeed, this trade-off depends on the distribution of endpoints within groups of each particular deployment, i.e. few groups with large amounts of endpoints vs. high number of groups with few endpoints each. Thus, it is not always the case that changing the endpoint's group is more scalable, but here  we present two examples form our experience:

\textbf{Acquisitions:} in case of an acquisition, the new employees are progressively moved through different groups until they get the same group as regular employees. The reverse also holds, when part of a company is sold, their users are moved to a group  that is associated with more restrictive policies. 

\textbf{Service insertion:} it is common that traffic has to go through middleboxes, e.g a firewall or a WAN optimizer. In some deployments, the SDA operators decide to update the group from the packets so that devices in the service chain decide whether to apply a policy or not. In other words, instead of applying different policies across the path for the same group, they change the group along the way so that different policies are applied across this same path.

\section{Related Work}

\subsection{SDN Pioneering Work}

Ethane \cite{casado2007ethane}, one of the first SDN designs, presents numerous similarities with our proposal: it also targets enterprise networks, presents a similar architecture with a centralized controller, and supports incremental deployment. However, Ethane specifically focuses on three key elements, mostly around the control plane: (i) providing network access control, (ii) a rich high-level policy language, and (iii) controller design and fault-tolerance. Conversely, in this paper we pay more attention data plane related aspects such as network isolation and resource efficiency.

Campus networks also motivated the inception of OpenFlow \cite{mckeown2008openflow}, that champions the decoupling of data plane and control plane, gives a strong focus to the  interface between the router and the controller and defines and approach to implementing rich policies on capable switches. SDA in its turn gives more emphasis to ability to support scalability in heterogeneous environments with devices with different capabilities.

Finally, SANE \cite{casado2006sane} offers a simple, high-level policy interface like the one used by SDA, but SANE tackles the problem in the border between L2 and L3 and does not trust the data plane routers. 

\subsection{BeyondCorp and Zero Trust Networks}\label{sec:beyondcorp}

In terms of securing the enterprise network, a closely related work is Beyond Corp \cite{beyondcorp}, also known as the Zero Trust model. Like SDA, Beyond Corp focuses on access control between network endpoints, with an especial emphasis on user-to-server connections. On the other hand, it should be noted that networking improvements (e.g. seamless mobility, etc) are not in the scope of BeyondCorp and therefore this section only discusses how SDA relates to BeyondCorp in terms of enterprise security.

BeyondCorp offers a solid approach to build secure enterprise networks by means of keeping healthy endpoints and redirecting traffic through access proxies. By doing so, BeyondCorp presents a security model that is agnostic to the underlying networking infrastructure. While this is a reasonable approach in certain scenarios, in our operational experience, we have found certain enterprise requirements that are hard to meet with a BeyondCorp-only approach. First, while BeyondCorp protects the access to the enterprise network, its focus is on protecting the access to enterprise applications at layer 7. However, it is not always possible to redirect traffic through the proxies (e.g. L2 traffic). Second, despite enterprise efforts around building healthy fleets of devices, the reality we observe is that insecure devices are still present in typical enterprise networks (even more due to the explosion of different IoT devices). These endpoints make the BeyondCorp approach harder to implement and, in many cases, secure on-boarding and admission still need to be performed by the network infrastructure. Third, with only a BeyondCorp security model the overall network performance could be degraded by malicious actors attempting to get access (even if unsuccessful) or explicitly looking to disrupt the network operation. SDA operates lower in the stack and not only protects the connection of devices to the network but also the network infrastructure itself. We believe that SDA complements BeyondCorp and the combination of both could contribute to strengthen the overall security of an enterprise network.

\subsection{Other Related Work}
Software Resolved Networks \cite{lebrun2018software} also leverages a reactive protocol (DNS extensions) and a centralized database, but requires involving endpoints when resolving routes, and allows a wider range of policies than our proposal.

Other proposals for enterprise networks center their work on specific elements of an enterprise network, such as ACL configurations \cite{aclUpdate}, systematic design of VLANs and ACL placement \cite{sung2011towards} or incremental SDN deployment \cite{levin2014panopticon}.

\section{Conclusion}
This paper presents SDA, a solution designed for modern enterprise networks. The main goal of the architecture is supporting emerging requirements, with a strong focus on mobility, segmentation, and incremental deployment. At the same time, it provides scalability and optimizes data plane resources for heterogeneous environments with devices of diverse capabilities.

SDA leverages common practices in networking (centralized control, network overlays) and makes use of a reactive approach to distribute network information and support mobility. Experimental results show that, when compared with traditional approaches, SDA exceeds a 70\% reduction in the overall forwarding state used by the network. Also, in very large deployments that need to deal with massive mobility events, network convergence is an order of magnitude faster than the status quo. For all these reasons we believe that SDA is a proper choice to address the unique requirements of modern enterprise networks that include scalable mobility, end-to-end segmentation, simplified administration, and overall resource optimization.

\section*{Acknowledgment}
The authors would like to thank Lorand Jakab for his invaluable help with the data collection scripts. Also Satish Kondalam regarding insights about customers, Satya Nanduri, Nirav Turakhia, and David Dai for the access to the deployments, Dipesh Patel, Ragupathi Rajavel, Raj Kumar, and Arash Albandi for the policy data, and Kevin Regan, Darrin Miller, and Mark Basinski for their help with the policy section. Thanks to Balaji Pitta Venkatachalapathy for the data collection, analysis and protocol tuning for the mobility use case. Thanks also to Pere Monclus for his valuable feedback on the manuscript, and Dino Farinacci for his LISP insights, inspiration, and vision. 

This work was partially supported by the Spanish MINECO under contract TEC2017-90034-C2-1-R (ALLIANCE) and the Catalan Institution for Research and Advanced Studies (ICREA).

\footnotesize
\bibliographystyle{unsrtnat}
\bibliography{en-biblio}

\end{document}